\documentclass[a4paper, onecolumn]{mn2e}
\usepackage{natbib_jrm,times, amsmath}
\usepackage{graphicx,pdflscape,amssymb}
\bibpunct[, ]{(}{)}{;}{a}{}{,}

\def \aj {AJ}
\def \mnras {MNRAS}
\def \pasp {PASP}
\def \apj {ApJ}
\def \apjs {ApJS}
\def \apjl {ApJL}
\def \aap {A\&A}
\def \nat {Nature}

\def \aaps {A\&A Suppl.}
\def \pasa {PASA}

\def\lesssim{\mathrel{\hbox{\rlap{\hbox{\lower4pt\hbox{$\sim$}}}\hbox{$<$}}}}
\def\gtrsim{\mathrel{\hbox{\rlap{\hbox{\lower4pt\hbox{$\sim$}}}\hbox{$>$}}}}

\long\def\symbolfootnote[#1]#2{\begingroup%
  \def\thefootnote{\fnsymbol{footnote}}\footnote[#1]{#2}\endgroup}

\begin{document}
\title[The site of SN 2007gr]{A high mass progenitor for the Type Ic Supernova 2007gr inferred from its environment}
\author[Maund \& Ramirez-Ruiz]{Justyn R. ~Maund$^{1}$\thanks{email: j.maund@sheffield.ac.uk}\thanks{Royal Society Research Fellow} \& Enrico Ramirez-Ruiz$^{2}$\\
$^{1}$ Department of Physics and Astronomy, University of Sheffield, Hicks Building, Hounsfield Road, Sheffield S3 7RH, U.K.\\
$^{2}$ Department of Astronomy \& Astrophysics, University of California, Santa Cruz, CA 95064, U.S.A.\\
}
\maketitle
\begin{abstract}
We present an analysis of late-time Hubble Space Telescope Wide Field Camera 3 and Wide Field Planetary Camera 2 observations of the site of the Type Ic SN 2007gr in NGC 1058.  The SN is barely recovered in the late-time WFPC2 observations, while a possible detection in the later WFC3 data is debatable.  These observations were used to conduct a multiwavelength study of the surrounding stellar population.  We fit spatial profiles to a nearby bright source that was previously proposed to be a host cluster.  We find that, rather than being an extended cluster, it is consistent with a single point-like object.  Fitting stellar models to the observed spectral energy distribution of this source, we conclude it is A1-A3 Yellow Supergiant, possibly corresponding to a star with $M_{ZAMS} = 40M_{\odot}$.
SN 2007gr is situated in a massive star association, with diameter of $\approx 300\,\mathrm{pc}$.  We present a Bayesian scheme to determine the properties of the surrounding massive star population, in conjunction with the Padova isochrones.  We find that the stellar population, as observed in either the WFC3 and WFPC2 observations, can be well fit by two age distributions with mean ages: $\sim 6.3\,$Myr and $\sim 50\,$Myr.  The stellar population is clearly dominated by the younger age solution (by factors of 3.5 and 5.7 from the WFPC2 and WFC3 observations, respectively), which corresponds to the lifetime of a star with $M_{ZAMS} \sim 30M_{\odot}$.  This is strong evidence in favour of the hypothesis that SN 2007gr arose from a massive progenitor star, possibly capable of becoming a Wolf-Rayet star.
\end{abstract}

\section{Introduction}
\label{sec:intro}
All stars with initial masses $>8M_{\odot}$ are expected to end their lives as
core-collapse Supernovae (SNe) or, possibly, disappearance due to direct infall onto black holes  during core-collapse \citep{2008ApJ...684.1336K}.
Recent efforts to detect SN
progenitors have been successful for Type IIP SNe, confirming that the progenitors of these particular SNe to be red supergiants (RSGs) with initial masses
$8<M_{init}<16M_{\odot}$ \citep{2008arXiv0809.0403S, 2009Sci...324..486M}.  The
progenitors of the H-deficient Type Ibc SNe have, however, proved difficult to detect.  Stellar evolution models predict that the
progenitors for these types of SNe are either very massive single
stars ($M_{init} \gtrsim 30-40M_{\odot}$), which undergo extreme
levels of mass loss during phases as Luminous Blue Variables and
Wolf-Rayet (WR) stars (removing their outer hydrogen layers), and lower
mass stars in binaries, which are stripped of their outer hydrogen layers by
mass transfer onto a binary companion \citep[see e.g.][]{eld04, 2013A&A...558A.131G,izzgrb,2004ApJ...612.1044P}.  The progenitors for these
types of SNe are expected to be blue, compact stars with high
effective temperatures.  Such stars lie on the blue-side of the
Hertzsprung-Russell (HR) diagram, whereas the RSG
progenitors of Type IIP SNe occupy the low effective temperature, red
side of the HR diagram.  

The review of pre-explosion observations of 12 Type Ibc SNe presented by \citet{2013MNRAS.436..774E} concluded that, statistically, it was unlikely that all could have arisen from single massive WR stars; suggesting a significant contribution to the Type Ibc SN rate from the binary progenitor channel.   For the study by  \citeauthor{2013MNRAS.436..774E} the case of SN~2002ap was particularly important, with the deep pre-explosion images effectively ruling out all massive single stars \citep{2007MNRAS.381..835C}.
Similar conclusions, based on the ratio of Type Ibc SN to H-rich SNe and the slope of the Initial Mass Function (IMF), suggested that Type Ibc SNe were too common to just arise from single massive stars \citep{2011MNRAS.412.1522S}.   More recently, the progenitor of the Type Ibc SN iPTF 13bvn has been authoritatively detected in pre-explosion observations.  While some analyses of the properties of the pre-explosion progenitor candidate suggest a possible single massive progenitor star, other studies of the pre-explosion source and models of the properties of the SN light curve are all consistent with a lower mass progenitor that was stripped through interaction with a binary companion  \citep{2013ApJ...775L...7C,2013A&A...558L...1G,2014A&A...565A.114F,2014AJ....148...68B,2015MNRAS.446.2689E}.  

There have been some suggestions of bright, massive progenitor stars detected in pre-explosion observations of interacting Type IIn SNe, such as 2005gl \citep{2007ApJ...656..372G, galyam05gl}, 2010jl \citep{2011ApJ...732...63S} and the possible SN 2009ip \citep{2011ApJ...732...32F}.  Limitations in the nature of the pre-explosion imaging for these SNe, however, make it difficult to draw firm conclusions about the nature of their progenitors.  Such Type IIn SNe are, however, rare and the fate of the majority of massive stars with $M_{init} > 20M_{\odot}$ remains uncertain.

Here we report the results of late-time observations  ($\sim 2.4$ years post-explosion) of the site of the Type Ic SN 2007gr in the galaxy NGC 1058 in an effort to gather further information about the nature of the progenitor star.  Pre-explosion observations of the site of SN 2007gr, in two Hubble Space Telescope (HST) Wide Field Planetary Camera 2 (WFPC2) bands, were reported by \citet{2008ApJ...672L..99C}.  \citeauthor{2008ApJ...672L..99C} did not find a source coincident with the SN position, but instead identified a nearby bright source as a possible host cluster and determined a mass estimate for the progenitor based on the age of such cluster in comparison with the lifetimes of possible progenitors.   The properties of this candidate host cluster were further analysed with additional HST WFPC2 observations by \citet{2014ApJ...790..120C}.
We adopt a distance modulus to NGC 1058 of $\mu = 30.13\pm0.35$ \citep{1994ApJ...432...42S}, which was also used in the previous analyses of \citet{2008ApJ...672L..99C} and \citet{2014ApJ...790..120C}, and an explosion epoch of 2007 Aug 13 (JD $2 454 325.5 \pm 2.5$; \citealt{2009A&A...508..371H}).  Applying the O3N2 metallicity measure  \citep{2004MNRAS.348L..59P} to the measured strengths of emission lines of {\sc H ii} regions in NGC 1058 \citep{1998AJ....116..673F}, we estimate the oxygen abundance at the deprojected radius of SN 2007gr of $R/R_{25} = 0.487$ to be approximately solar \citep{2004A&A...417..751A}.
\section{Observations and Data Reduction}
\label{sec:obs}
A log of the pre-explosion, post-explosion and late-time HST observations of the site of SN~2007gr is presented in Table \ref{tab:obs:log}.   The HST WFPC2 observations were retrieved from the STScI archive\footnote{http://archive.stsci.edu}, having been processed through the On-The-Fly Recalibration pipeline.  Pairs of exposures for each filter, at each epoch, were combined, for display purposes, using the {\sc PyRAF}\footnote{STSDAS and PyRAF are products of the Space Telescope Science Institute, which is operated by AURA for NASA} task {\it astrodrizzle}.   Photometry of the WFPC2 frames was conducted using the {\sc dolphot} package\footnote{http://americano.dolphinsim.com/dolphot/} \citep{dolphhstphot} with the specific WFPC2 module.  The {\sc iraf} task {\it crrej} was used to create a cosmic ray mask, which enabled the removal of cosmic rays from the pairs of exposures for each filter prior to the application of {\sc dolphot}.  For the post-explosion ($\sim 78$d post-explosion) and late-time ($\sim 465$d post-explosion) WFPC2 observations, the SN was imaged on the Planetary Camera (PC) chip, with pixel scale 0.046 arcsecs, while for the pre-explosion observations ($\sim 2231$d pre-explosion) the SN position fell on the Wide Field 2 (WF2) chip with pixel scale 0.0996 arcsec.

Later WFC3/UVIS observations ($\sim 882$d post-explosion) of the site of SN 2007gr were acquired using a 4-point box dither pattern for each of the three filters used.  The individual distorted images were retrieved from the STScI archive and combined and corrected for geometric distortion using the {\it astrodrizzle} package.  Due to the nature of the dither pattern, it was possible to resample the image to a finer pixel scale of 0.02 arcsecs (compared to the original WFC3/UVIS pixel scale of 0.04 arcsecs).  Photometry of the WFC3 frames was also conducted using {\sc dolphot}, with the WFC3 specific module, using the drizzled, distortion corrected images as references.

In both the late-time WFPC2 and WFC3 frames, detection limits for those stars not detected at certain wavelengths were established through artificial star tests.  Artificial stars were inserted at the positions determined for the stars at wavelengths in which the stars were detected.   The completeness of the recovery of the artificial stars (at $\mathrm{S/N = 3}$, within 1 pixel of the inserted position) as a function of input magnitude was modelled by a cumulative normal distribution, parameterised by the magnitude at which 50\% completeness was achieved.   Given the complexity of the region hosting SN 2007gr, the derived detection thresholds were found to vary significantly across the field due to varying background levels and stellar crowding; therefore, detection thresholds were calculated independently for each star.  Artificial star tests at random locations around the position of SN 2007gr were also used to calculate corrections to the uncertainty on the photometry due to crowding effects.

SN 2007gr was still bright in the post-explosion WFPC2 observations, and the position of the SN in the F450W image, relative to surrounding, nearby stars, was used to determine the SN position in the pre-explosion WFPC2 and late-time WFPC2 and WFC3 observations using the {\sc iraf} task {\it geomap}.  The post-explosion WFPC2 F450W image was used to determine the SN position in the pre-explosion WFPC2 F450W observation with an accuracy of 37 milliarcsecs.  Geometric transformations were determined between the post-explosion WFPC2 F450W image and the late-time WFPC2 and WFC3 F555W observations.  The SN position was identified on the late-time WFPC2 and WFC3 images to within 15 and 11 milliarcsecs, respectively.  Small shifts between the late-time WFPC2 and WFC3 observations taken with other filters were calculated using the {\sc PyRAF} task {\it crosscor} and were generally found to be small ($\sim 1$pix), with associated uncertainties that were significantly smaller.

\begin{table*}
\caption{\label{tab:obs:log} HST observations of the site of SN~2007gr}
\begin{tabular}{lccccl}
\hline\hline
Date                & Instrument   &  Filter & Exposure   &   Pixel     \\
                         &     &               & Time (s)   & Scale ($\arcsec$)\\
\hline \multicolumn{5}{c}{{\bf Pre-explosion}}\\
2001 Jul 3.72$^{1}$& WFPC2/WF2 & F450W & 460 & 0.097 \\
2001 Jul 3.74$^{1}$ &WFPC2/WF2 & F814W & 460 &0.097\\
\\ 
\multicolumn{5}{c}{{\bf Post-explosion}}\\
2007 Oct 29.78$^{2}$ & WFPC2/PC1 &F450W & 700 & 0.046 \\
2007 Oct 29.80$^{2}$ & WFPC2/PC1 &F814W & 320 & 0.046 \\
\\
\multicolumn{5}{c}{{\bf Late-time}}\\
2008 Nov 20.66$^{3}$& WFPC2/PC1 & F555W & 460 & 0.046\\
2008 Nov 20.67$^{3}$ & WFPC2/PC1 & F814W & 700 & 0.046 \\
\\
2008 Nov 25.59$^{3}$ & WFPC2/PC1 & F450W & 800 & 0.046\\
2008 Nov 25.60$^{3}$ & WFPC2/PC1 & F675W & 360 & 0.046 \\
\\
2010 Jan 10.81$^{4}$ & WFC3/UVIS & F336W & 1060 & 0.02 \\
2010 Jan 10.83$^{4}$ & WFC3/UVIS & F555W & 640 & 0.02\\
2010 Jan 10.38$^{4}$ & WFC3/UVIS & F625W & 640 &0.02 \\
\hline\hline
\end{tabular}
\\
$^{1}$ SNAP-9042 (PI: S. Smartt)\\
$^{2}$ GO-11119 (PI: S. Van Dyk)\\
$^{3}$ SNAP-10877 (PI: W. Li)\\
$^{4}$ GO-11675 (PI: J.R. Maund)\\
\end{table*}

\section{Results and Analysis}
\label{sec:res}
\begin{figure*}
\includegraphics[height=6.5cm]{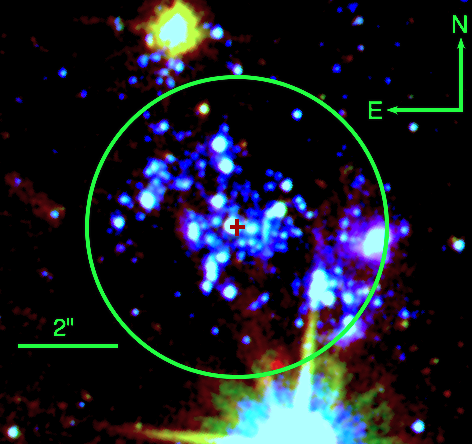}
\caption{A three colour HST WFC3/UVIS image of the site of SN 2007gr, with the position of the SN marked by the red cross (at the centre of the field).  The circle indicates the approximate diameter of the young, massive star population around the SN position of $6\arcsec$.  The two bright sources to the North and South of the SN position are foreground stars.}
\label{fig:res:image}
\end{figure*}
\begin{figure*}
\includegraphics[height=6.5cm]{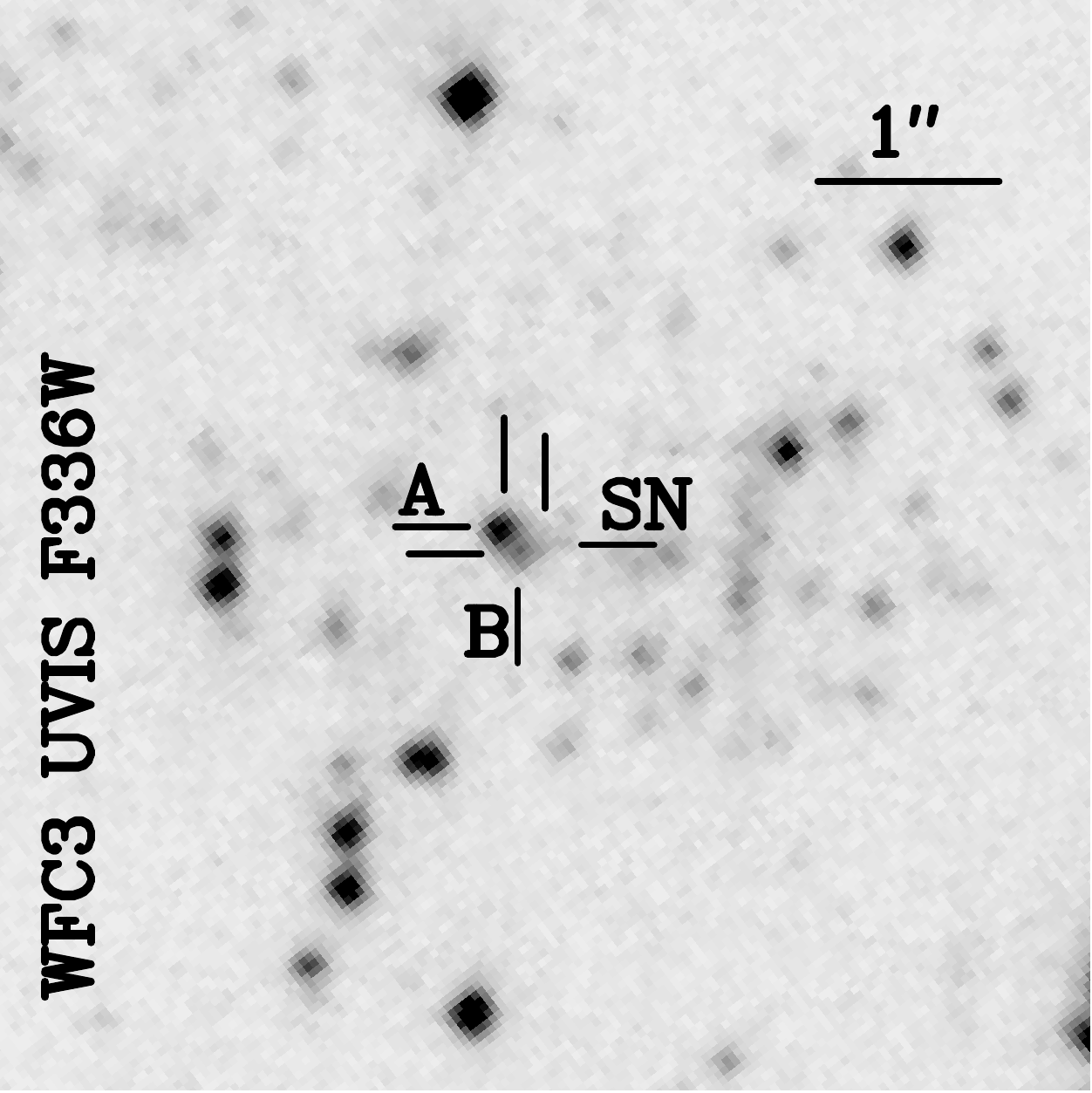}
\includegraphics[height=6.5cm]{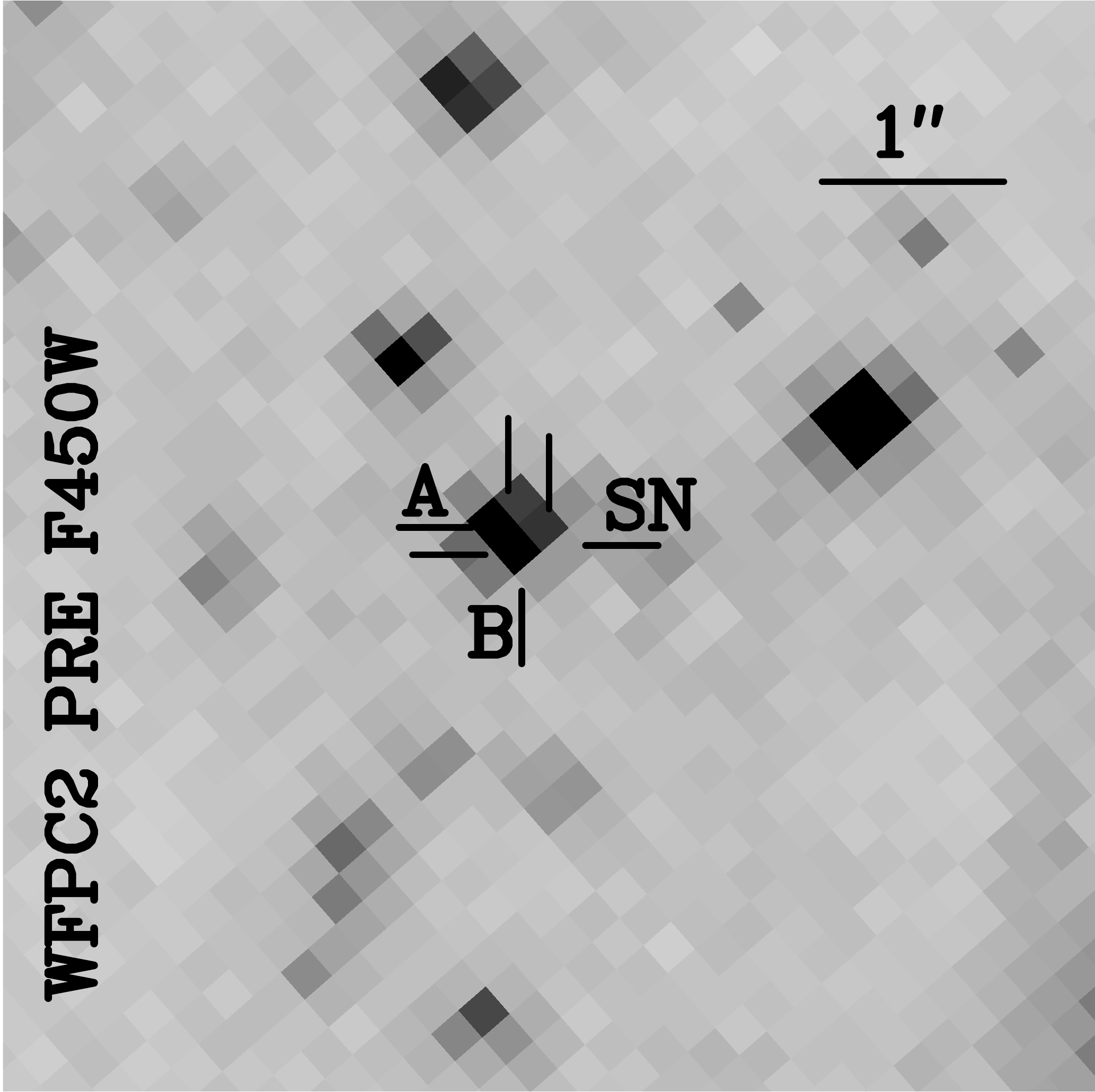}
\caption{ The site of SN~2007gr observed with HST. {\it Left)}  A zoom-in on the site of SN~2007gr as imaged by WFC3 on 2010 Jan 10 using the F336W filter.  No source is recovered at the SN position.  Two nearby stars, labelled A and B, are located close to, but not coincident, with the SN position. {\it Right)} The site of SN~2007gr observed before explosion with the HST WFPC2 WF2 chip using the F450W filter.}
\label{fig:res:latepre}
\end{figure*}
\subsection{The location of SN~2007gr}
\label{sec:res:location}
The site of SN~2007gr, in the late-time observations, is presented in Figures \ref{fig:res:image} and \ref{fig:res:latepre}.  The SN resides in a bright, young stellar population dominated by blue stars covering a region of approximately 6 arcsecs in diameter.  Our analysis of the pre-explosion WFPC2 observations confirms the results of \citet{2008ApJ...672L..99C} and \citet{2014ApJ...790..120C}, who similarly found there was no source at the SN position in both the pre-explosion F450W and F814W images.  Using {\sc dolphot}, we estimate pre-explosion detection limits at the SN position of  $m_{F450W} = 22.1\pm1.0$ and $m_{F814W} = 22.6\pm1.0$ mags.

In the late-time WFPC2 observations we recover the fading SN, with brightness $m_{F450W}=22.19 \pm 0.03$, $m_{F555W} = 21.69 \pm 0.02$, $m_{F675W}=21.36 \pm 0.03$ and $m_{F814W}=21.32 \pm 0.02$ mags.   The SN is clearly not detected in the late-time WFC3 F336W image, however {\sc dolphot} suggests a source might have been recovered at the SN position in the corresponding F555W and F625W images. We note, however, that these detections are not confirmed in each of the four constituent dithered exposures and are complicated by the location of the SN position relative to the wings of the Point Spread Function (PSF) of a nearby star (see Section \ref{sec:res:srcA}), that is particularly bright at these wavelengths, and complexities in the WFC3/UVIS PSF in these filters.  We tested these possible detections by inserting faint artificial stars at the SN position, with a view to simulating different noise conditions at that location.  We found that the reported detection was highly sensitive to even the slightest modifications of the pixel counts at and around the SN position.  This suggests that the reported detections, if real, would be at the detection limit at the SN position and any photometry would be extremely uncertain, being highly sensitive to readout noise and noise associated with flux originating from the bright nearby star.  As such, we cannot confidently report the detection of SN 2007gr or any other object at the SN position in the late-time WFC3 observations.   For the WFC3 observations, we determined detection limits of $m_{F336W} = 24.8\pm 0.2$, $m_{F555W}=25.3 \pm 0.1$ and $m_{F625W}= 24.8\pm 0.1$ mags.
\subsection{Nearby Sources}
\label{sec:res:srcA}
In both the late-time WFPC2 and WFC3 observations we recover a bright source in close proximity (offset by 0.12 arcsecs East) to the position of SN~2007gr, previously identified by \citet{2008ApJ...672L..99C}, and labelled source A (see Fig. \ref{fig:res:latepre}).  In the late-time WFC3/UVIS F336W image we also recover another, fainter source at $m_{F336W} = 22.87 \pm 0.04$ mags, which is offset from the SN position by 0.06 arcsecs, approximately South East (which we label Source B).  Source B was not detected in any of the other WFC3 bands, suggesting that this is a blue source that, if it is star, has a spectral type of B2 or earlier.\\

Photometry of Source A in the late-time WFC3/UVIS and WFPC2 images is presented in Table \ref{tab:res:srcAphot}.   We note that the late-time brightness of Source A is consistent with the brightness of the source measured in the pre-explosion WFPC2 F450W and F814W observations, suggesting that Source A is not variable.  Previously, \citet{2008ApJ...672L..99C} and \citet{2014ApJ...790..120C} presented interpretations of the Spectral Energy Distribution (SED) of Source A in the context of it being a compact cluster, that might be related to the progenitor (despite the large offset corresponding to a projected distance of $6.2\,\mathrm{pc}$).  The analyses of \citeauthor{2008ApJ...672L..99C} and \citeauthor{2014ApJ...790..120C} were based on subsampled WFPC2  observations from 2001 and 2008, respectively.  The late-time WFC3 observations were purposefully acquired to resample the PSF to probe the spatial extent of source A and determine if it was a star or a compact cluster.\\
The {\sc dolphot} photometry package returns measures of the shape and apparent physical extent of sources, through the $\chi^{2}$ and sharpness parameters.  For all the late-time WFC3 and WFPC2 images, {\sc dolphot} identified Source A as a point-like source.  We also utilised the {\sc ishape} package \citep{1999A&AS..139..393L}\footnote{http://baolab.astroduo.org/} to fit the observed spatial profile of Source A with underlying cluster profiles (here we used elliptical  Moffat profiles of order 1.5) convolved with the corresponding PSF for each resampled WFC3 image.  Two iterations of the {\sc ishape} analysis were conducted with empirical PSFs for each of the drizzled, resampled WFC3 images, derived directly from the images using the {\sc daophot} package running under {\sc IRAF}, and PSFs computed using the TinyTim PSF simulator\footnote{http://www.stsci.edu/software/tinytim/tinytim.html}.  The results from these analyses were found to be identical.  To assess the suitability of an extended, cluster-like profile to explain the observed spatial properties of Source A, we compared the quality of the fits using the Moffat function with fits derived using a simple delta function convolved with the appropriate PSF (corresponding to a simple stellar-like, point source).  Following the analysis of \citet{2007A&A...469..925S}, we adopt the ratio of the $\chi^{2}$-values for fits using the two underlying functions to assess the relative merit of the extended cluster-like profile.  The results of the fits applied to Source A, as observed in the late-time WFC3 F336W, F555W and F625W images, are presented in Table \ref{tab:res:ishape}.  We find that the extended Moffat profile does not provide a significantly better fit to the spatial profile of Source A than a simple delta function convolved with the PSF.  We also inspected the output images produced by {\sc ishape}, in which the model fit to Source A was subtracted, and find that the only appreciable difference between the Moffat and delta function fits was a slight modification to the background count levels.  We converted the corresponding values of the  Full Width at Half Maximum (FWHM) for the Moffat profiles, modified for the ellipticity of the profile, to values of effective radius ($R_{eff}$; \citeauthor[see][and their Equation 2]{2007A&A...469..925S}).  Given the scales down to which {\sc ishape} is sensitive to truly extended profiles \citep[FWHM $\sim$ 0.2px][]{2004A&A...416..537L}, we find that the values of the effective radius returned by {\sc ishape} are below this threshold (corresponding to 0.2 pc at the distance of NGC 1058) for Source A as observed in the F555W and F625W images.  The inferred FWHM in the F336W image is only slightly larger than this threshold, however we note that the S/N level for Source A is significantly lower than for the other filters and  that the quality of the fit is also affected by the proximity of the nearby Source B.  The values of the effective radius of Source A, if it is extended, are also significantly smaller than the limit ($R_{eff} \sim 1.2\mathrm{pc}$) for the cluster sample, for example, observed by \citet{2005A&A...431..905B}  in M51.  We find no evidence, therefore, for Source A being an extended source.\\

\begin{figure*}
\includegraphics[width=8cm,angle=270]{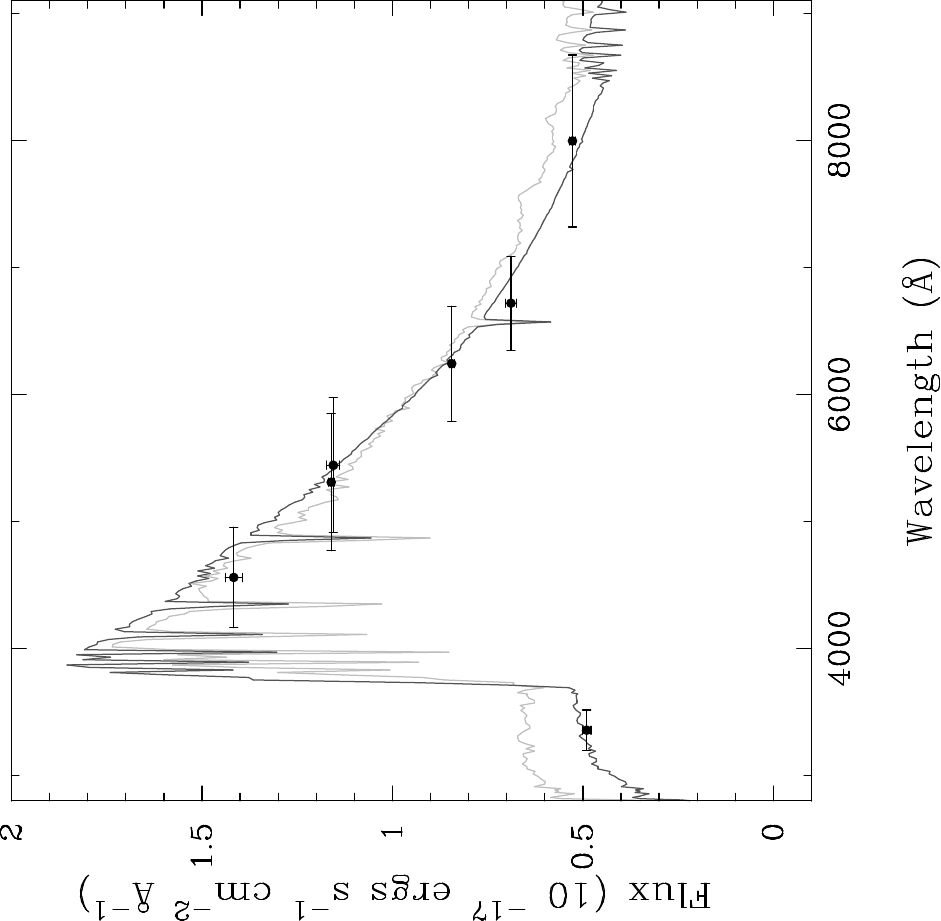}
\caption{The observed spectral energy distribution of Source A from late-time WFC3/UVIS and WFPC2 photometry.  Overlaid are the best fit theoretical SEDs for clusters (Starburst99 with $\tau =  8.48$; light grey) and supergiant stars (ATLAS9 with $T_{eff} = 9000K$; dark grey).}
\label{fig:res:srcASED}
\end{figure*}

We also examined the observed SED of Source A, in the three WFC3 UVIS and four WFPC2 bands using our own Bayesian SED fitting algorithm, in comparison with ATLAS9 \citep{2004astro.ph..5087C} stellar and STARBURST99 \citep{1999ApJS..123....3L} cluster SEDs.  For each fit we assumed a standard \citet{ccm89} $R_{V} = 3.1$ reddening law and varied the effective temperature and age for the stellar and cluster SEDs, respectively. The results of these fits are presented in Figure \ref{fig:res:srcASED}.   Comparing the Bayesian evidence computed for each of the fits, we calculate a Bayes factor of $5 \times 10^{63}$ which overwhelmingly supports the hypothesis that the underlying SED that describes the photometry of Source A is a stellar SED.  This result is primarily driven by the difference in the predicted F336W magnitude and the size of the Balmer jump between the stellar and cluster models.\\  

\begin{figure*}
\includegraphics[width=8cm,angle=270]{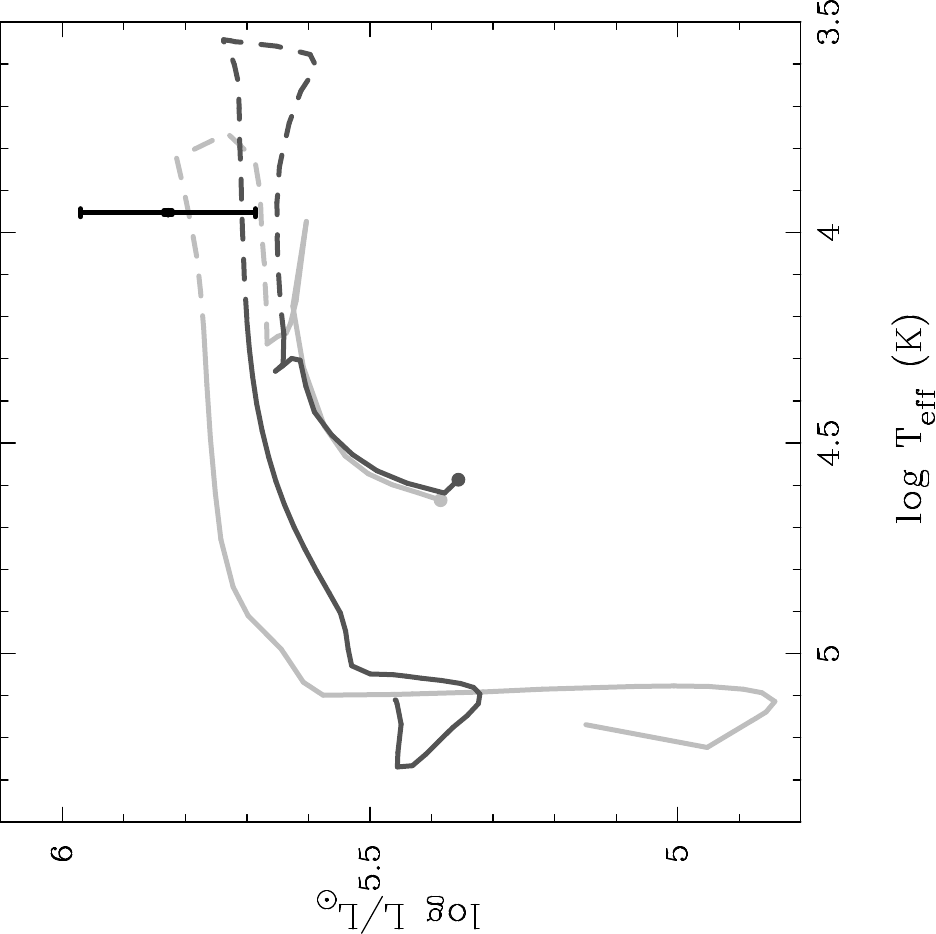}
\caption{The location of Source A on the HR diagram.  Overlaid are stellar evolution tracks from the Cambridge STARS \citep[Dark grey;][]{eld04} and the PARSEC \citep[Light grey;][]{2012MNRAS.427..127B, 2013EPJWC..4303001B} models for a star with $M_{ZAMS} = 40M_{\odot}$.  The dashed portions of each track correspond to brief phases of evolution corresponding to $\sim 10^{4}$ and $10^{5}$ years, of a total lifetime of 5.1 Myr, for the PARSEC and STARS models, respectively.}
\label{fig:res:srcAHRD}
\end{figure*}

From the considerations of both the observed SED and the spatial extent of Source A, we conclude that this object is in a fact a star and not a cluster, as previously assumed.  The best fit parameters for the stellar fit to the SED of Source A correspond to $T_{eff} = 8970 \pm 85 \mathrm{K}$ and $E(B-V) = 0.3 \pm 0.01$ mags, or an A1-3 supergiant following the calibration of \citet{schmidtkaler} and its application to the ATLAS9 models presented in the {\sc synphot} manual\footnote{http://www.stsci.edu/institute/software\_hardware/stsdas/synphot/SynphotManual.pdf}.  The inferred reddening is higher than the value of $E(B-V) = 0.092$ estimated by \citet{2008ApJ...673L.155V} and \citet{2010MNRAS.408...87M} and the similar value derived by \citet{2014ApJ...790..120C} from measurements of the interstellar Na {\sc i} line strength for SN~2007gr.  The position of Source A on the HR diagram is shown on Figure \ref{fig:res:srcAHRD}, and the inferred luminosity of Source A is found to be consistent with a $40M_{\odot}$ star undergoing a short phase of evolution as a Yellow Supergiant \citep{eld04}.

\begin{table*}
\caption{\label{tab:res:srcAphot} Photometry of Source A in the late-time HST WFC3/UVIS and WFPC2 observations.}
\begin{tabular}{cccccccc}
\hline\hline
\multicolumn{3}{c}{WFC3/UVIS} & & \multicolumn{4}{c}{WFPC2} \\
\cline{1-3}\cline{5-8}\\
$F336W$ & $F555W$ & $F625W$ & &$F450W$ & $F555W$ & $F675W$ & $F814W$ \\
\hline
22.057 & 21.333 & 21.152 & & 21.512 & 21.259 & 21.121 & 20.857 \\
(0.025) & (0.007)  & (0.008) & & (0.017) & (0.016) & (0.022) & (0.017) \\
\hline\hline
\end{tabular}
\end{table*}
\begin{table*}
\caption{\label{tab:res:ishape}Results from ISHAPE fits to Source A in the WFC3 UVIS images}
\begin{tabular}{ccc}
\hline\hline \\
           &  $\chi^{2}_{\mathrm{Moffat}}/\chi^{2}_{\mathrm{Delta}}$ & $R_{eff}\,$ (pc) \\
\hline\\
F336W & 0.98 & 0.42 \\
F555W & 0.99 & 0.10 \\
F625W & 0.99 & 0.10 \\
\hline\hline
\end{tabular}
\end{table*}

\subsection{The Surrounding Stellar Population}
As noted previously, SN 2007gr is located at the centre of a young, massive star complex of approximate size  300 pc in diameter (see Fig. \ref{fig:res:image}) and in close proximity to a $40M_{\odot}$ star.  The strong association with very massive stars, including those that might undergo a Wolf-Rayet phase, may have implications for the progenitor of SN 2007gr, if the progenitor star belonged to this population.  To measure the properties of the surrounding stellar population we adopt a Bayesian approach, based on the method \citet{2005A&A...436..127J} and other similar methods, such as that presented by \citet{2006ApJ...645.1436V}.\\
For a general star $i$, with observed photometric data $\mathbf{D}_{i}$, we are primarily interested in determining the posterior probability for the age ($\tau$ or $\log \left(t\, \mathrm{years}\right)$) and extinction ($A_{V}$) through comparison with predicted brightnesses and colours derived from stellar evolution models and presented as isochrones.  For this study we use the Padova isochrones\footnote{http://stev.oapd.inaf.it/cgi-bin/cmd} \citep{2002A&A...391..195G}.  For the comparison of the observed data with isochrones, however, we also depend on other parameters such as the assumed distance modulus ($\mu^{\prime}$) and the mass of the star ($M_{i}$).  In addition, one might also consider the effect of metallicity, however we will assume a single value corresponding to solar metallicity (see Section \ref{sec:intro}).\\
Following Bayes' Theorem, the posterior probability distribution can be expressed in terms of the likelihood and the prior probability of the vector of parameters that underly the isochrone model:
\begin{equation}
p(\tau, A_{V},\mu^{\prime}, M_{i} | \mathbf{D}_{i})  \propto    p(\mathbf{D}_{i}| \tau, A_{V},\mu^{\prime}, M_{i})  p(\tau, A_{V}, \mu^{\prime}, M_{i})
\end{equation}
\begin{equation}
p(\tau, A_{V} | \mathbf{D}_{i})  \propto   \int\int p(\mathbf{D}_{i}| \tau, A_{V},\mu^{\prime}, M_{i})  p(\tau, A_{V}, \mu^{\prime}, M_{i})\mathrm{d}\mu^{\prime}\mathrm{d}M_{i}
\label{eqn:post}
\end{equation}
where $ p(\tau, A_{V}, \mu^{\prime}, M_{i})$ constitutes the prior probabilities on the parameters, which we assume are independent.  Following \citet{2005A&A...436..127J}, we take the prior probability for the initial mass of the star to follow the form of the Initial Mass Function (IMF): 
\begin{equation}
p(M_{i}) = \frac{M_{i}^{-\alpha}}{\int^{M_{max}}_{M_{min}}M_{i}^{-\alpha}\mathrm{d}M_{i}}
\end{equation}
where the denominator normalises the prior over the range of masses present in a given isochrone and for a \citet{1955ApJ...121..161S} IMF $\alpha = 2.35$.  We note that $p(M_{i}) > 0$ only if $M_{min}(\tau) \le M_{i} \le M_{max}(\tau)$, where $M_{min}(\tau)$ and $M_{max}(\tau)$ are the minimum and maximum masses for an isochrone with age $\tau$.  The distance modulus $\mu^{\prime}$ is  a nuisance parameter, which serves to shift the isochrones relative to the observed data.  Given a reported distance modulus to the host galaxy of $\mu \pm \sigma_{\mu}$,  the prior probability for $\mu^{\prime}$ is $p(\mu^{\prime}) \sim N(\mu, \sigma^{2}_{\mu})$.   The initial mass $M_{i}$ serves as a coordinate along the length of the isochrone and for a given isochrone, the observed data may suggest a range of masses.  The mass is treated as a nuisance parameter and is also marginalised over.\\
The likelihood function is defined with the observed photometry in each filter compared directly to the predicted magnitudes from the isochrones.  Rather than comparing the observed photometry with isochrones in the colour-magnitude plane,  our approach specifically avoids complications associated with correlated uncertainties that arise in using colours, but necessitates the explicit use of the distance modulus as a nuisance parameter \citep{2009MNRAS.399..699M}.  The likelihood takes the form:
\begin{equation}
p(\mathbf{D}_{i} | \tau, A_{V}, M, \mu^{\prime}) =  \prod_{j} \frac{1}{\sqrt{2\pi}\sigma_{j}} 
\exp\left\{
-\frac{1}{2}\left(
\frac{m_j - m_j(\tau, A_{V}, M_{i}) - \mu^{\prime}}{\sigma_{j}}
\right)^{2}\right\}
\label{eqn:like}
\end{equation}
where $m_{j}$ is the $j$th {\it apparent} magnitude for star $i$ and $m_{j}(\tau, A_{V}, M_{i})$ is the {\it absolute} magnitude predicted for a star with mass $M_{i}$ lying on an ischrone with age $\tau$ and extinction $A_{V}$.  In addition, if the star is not detected in a given filter the corresponding detection limit ($m^{lim}_{j}$, $\sigma^{lim}_{j}$; see Section \ref{sec:obs}) can be substituted into the likelihood expression as:
\begin{equation}
p(m^{lim}_{j}, \sigma^{lim}_{j} |  \tau, A_{V}, M_{i}, \mu^{\prime}) = \frac{1}{2} \left\{ 1 + \mathrm{erf} \left(
\frac{m^{lim}_{j} - m_j(\tau, A_{V}, M_{i}) - \mu^{\prime}}{\sqrt{2}\sigma^{lim}_{j}} \right) \right\}
\end{equation}
This enables us to consider photometry covering a wide wavelength range that may be partially incomplete.  Following 
\citet{2005A&A...436..127J} we enforce the normalisation requirement for the likelihood such that:
\begin{equation}
 \int\int p(\mathbf{D}_{i} | \tau, A_{V})\mathrm{d}\tau\mathrm{d}A_{V} = 1
 \end{equation}
for flat priors on $\tau$ and $A_{V}$, where: 
\begin{equation}
p(\mathbf{D}_{i} | \tau, A_{V}) = \int\int p(\mathbf{D}_{i} | \tau, A_{V}, M_{i}, \mu^{\prime})p(M_{i}, \mu^{\prime})\mathrm{d}M_{i}{d}\mu^{\prime}
\end{equation}
\\
The Padova isochrones are only computed to reproduce the properties of single stars.  This is likely to provide an incomplete representation of the observed properties of a given stellar population, if a proportion of the observed sources are unresolved binaries with colours that deviate from single star isochrones due to the composite nature of their SEDs.  In order to consider the role of binarity on the observed stellar population, we adopt the approximation that the observed brightness and colours of a binary system arise from the simple addition of flux from two single stars that lie on the same isochrone.  This approximation is, however, only valid for non-interacting binaries, for which the evolution of the two stars proceeds independently, which in some cases may have significant flaws \citep{2012Sci...337..444S}.  We assign the mass $M_{i}$ to describe the mass of the primary star, and introduce an additional nuisance parameter $q$ (where $0 < q \leq 1$) to define the binary mass ratio (for which we assume a flat prior).  For each point along the isochrone we consider the magnitudes resulting from the sum of fluxes from a star with mass $M_{i}$ at that point and a secondary with mass equal or less than $M_{i}$.  In the equivalent form of Equation \ref{eqn:post}, for the binary star scenario the parameter $q$ is also marginilised over.  While a single star isochrone defines an infinitely narrow track across the magnitude-magnitude plane, the inclusion of binaries makes the isochrone "fuzzier", with brighter stars also accommodated by fainter single star isochrones.\\
The benefit of a Bayesian approach to this problem is that we do not need to explicitly determine if a star is a single star or a binary.  Bayesian mixture models enable us to embrace the ambiguity between the two cases, by considering the final probability as a sum of the single and binary scenarios:
\begin{equation}
p(\tau, A_{V} | \mathbf{D}_{i}) = \left( 1 - P_{bin}\right)p(\tau, A_{V} | \mathbf{D}_{i}, {\rm single}) + P_{bin}p(\tau, A_{V} | \mathbf{D}_{i},{\rm binary})
\end{equation}
where $P_{bin}$ is the probability of a source being a binary (or binary fraction), for which we adopt $P_{bin} = 0.5$.\\
Due to the non-monotonic nature of the predicted properties of isochrones and degeneracies in the underlying parameters, it is unlikely that there will be a unique solution for a given star in the $\tau-A_{V}$ plane.  In addition, if one considers a population of stars, there is no reason that this population may itself be described by a single, unique solution.  This issue becomes more problematic if the constituent stars arise from a spread of population characteristics.  In the Bayesian context it is convenient to consider this as an hierarchical problem, with the aim to identify the parameters of the underlying distributions from which the observed population of stars is drawn.  This approach uses each star as a separate probe of the underlying properties of the population, and uses the entire ensemble of stellar properties to resolve the degeneracies in the isochrone fitting problem.  Similarly to \citet{2013MNRAS.435.2171W}, we consider underlying age distributions of stars to take the form $N(\tau^{\prime}, \sigma^{\prime, 2}_{\tau})$, which have normal distributions in log(years) and log-normal distributions in years.  A given star may have age $\tau$, which is drawn from distributions parameterised by $\tau^{\prime}$ and $\sigma^{\prime 2}_{\tau}$.  Using Bayes' Theorem, the hierarchical form becomes:
\begin{equation}
p(\tau^{\prime}, \sigma^{\prime}_{\tau} | \mathbf{D}_{i}) \propto \int\int p(\mathbf{D}_{i} | \tau, A_{V})p(\tau | \tau^{\prime}, \sigma^{\prime}_{\tau}) p(\tau^{\prime}, \sigma^{\prime }_{\tau})p(A_{V})\mathrm{d}\tau\mathrm{d}A_{V}
\label{eqn:hierarchical}
\end{equation}
As shown by  \citet{2013MNRAS.435.2171W}, we may also consider the observed population as arising from a mixture of underlying distributions (or epochs of star formation), in analogy to the mixture of single and binary stars discussed above.  The mixture of underlying distributions can be described by the vector of parameters $\{\underline{\tau^{\prime}}, \underline{\sigma^{\prime}}, \underline{w} \}$, where $\underline{w}$ is a vector containing the relative weights of each component distribution in the mixture, such that $\sum_{k}w_{k} = 1$.  The likelihood of the observed data for the $i^{\mathrm{th}}$ star arising from the $k^{\mathrm{th}}$ distribution of ages is $w_{k}p(\mathbf{D}_{i} | \tau_{k}, \sigma_{k}, A_{V})$, such that for the entire set of observations $\mathbf{\underline{D}}$ for all $N_{star}$ stars:
\begin{equation}
p(\mathbf{\underline{\tau^{\prime}}}, \mathbf{\underline{\sigma^{\prime}_{\tau}}}, A_{V} | \mathbf{\underline{D}} ) \propto \prod^{N_{star}}_{i}\sum^{N_{m}}_{k}w_{k}p(\mathbf{D}_{i} | \tau_{k}, \sigma_{k}, A_{V})p(\tau_{k}, \sigma_{k}, A_{V})
\end{equation}
where the total number of components in the mixture is $N_{m}$ and it is assumed, in this case, that all populations share the same extinction $A_{V}$.  \citet{2013MNRAS.435.2171W} used a Reversal Jump Markov Chain Monte Carlo technique to determine $N_{m}$ as part of the fitting process.  We adopt a different approach by using the Nested Sampling algorithm \citep{2004AIPC..735..395S,2008MNRAS.384..449F},  with importance sampling \citep{2013arXiv1306.2144F}\footnote{http://ccpforge.cse.rl.ac.uk/gf/project/multinest/}, to calculate the Bayesian evidence or marginal likelihood ($Z$; the constant of proportionality in  Equation \ref{eqn:hierarchical}) for different $N_{m}$ and using Bayesian model selection to determine the optimal value of $N_{m}$ to describe the observed population.  In using Nested Sampling there are two important considerations:
\begin{enumerate}
\item{The labels for each underlying distribution are interchangeable and, as such, the Nested Sampling algorithm simultaneously finds all labelling permutations for $N_{m} > 1$.  This leads to the evidence $Z$ being overestimated by a factor of $N_{m}!$.  Due to the trivial nature of the correction that needs to be applied to the Bayesian evidence, it is convenient to find all modes, including mirror modes, and only sort the equally weighted posterior samples once the algorithm has converged, to find individual solutions for each component}
\item{The addition of mixture components in excess of the optimal number of components will not result in an increase in $Z$ (following the interpretation of \citealt{Jeffreys61}).  The minimum value of $N_{m}$ that maximises $Z/N_{m}!$ represents the most parsimonious solution, that adequately describes the observed population using the minimum number of parameters.}
\end{enumerate}

\begin{figure*}
\includegraphics[height=8cm, angle=270]{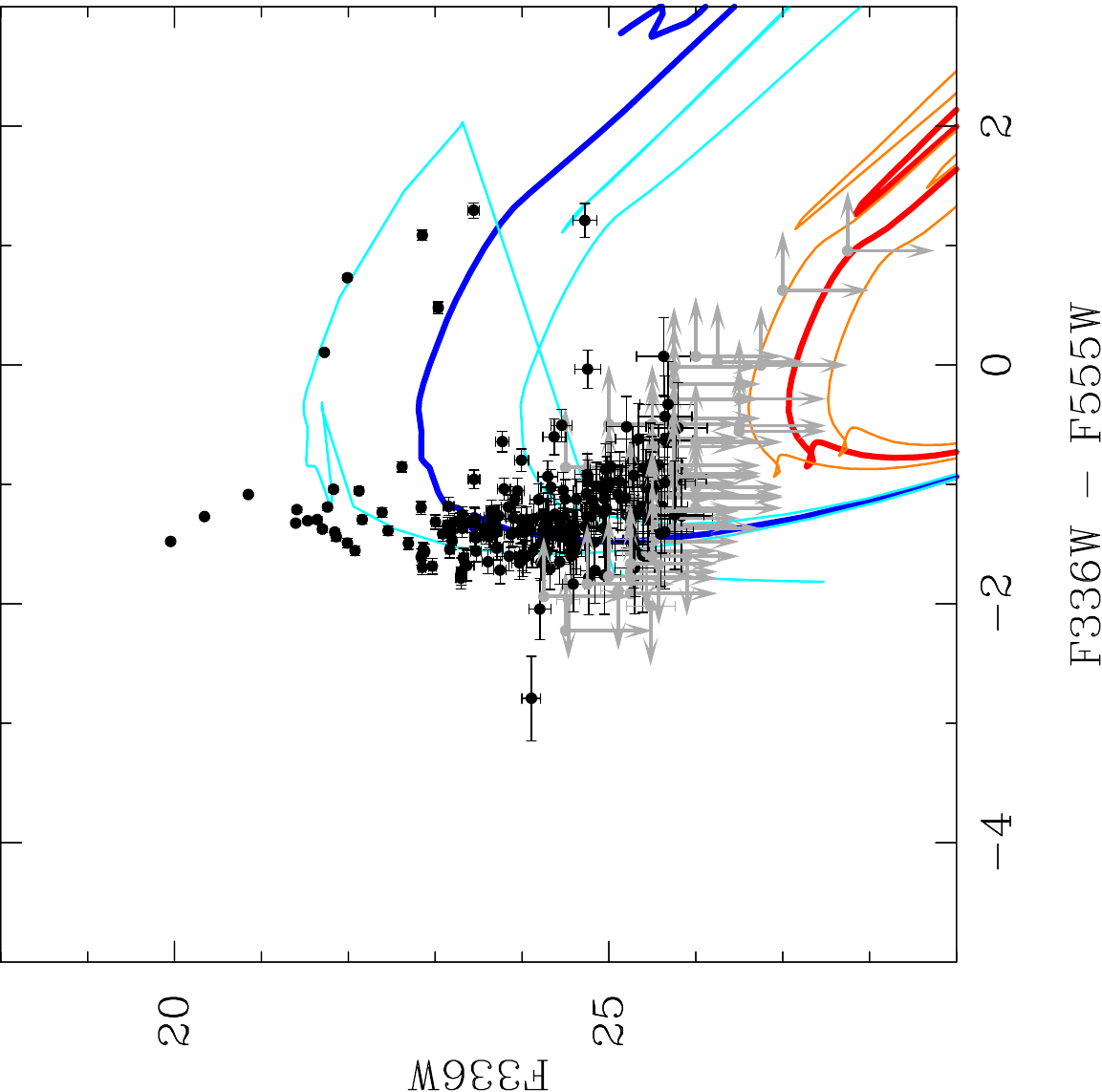}
\includegraphics[height=8cm, angle=270]{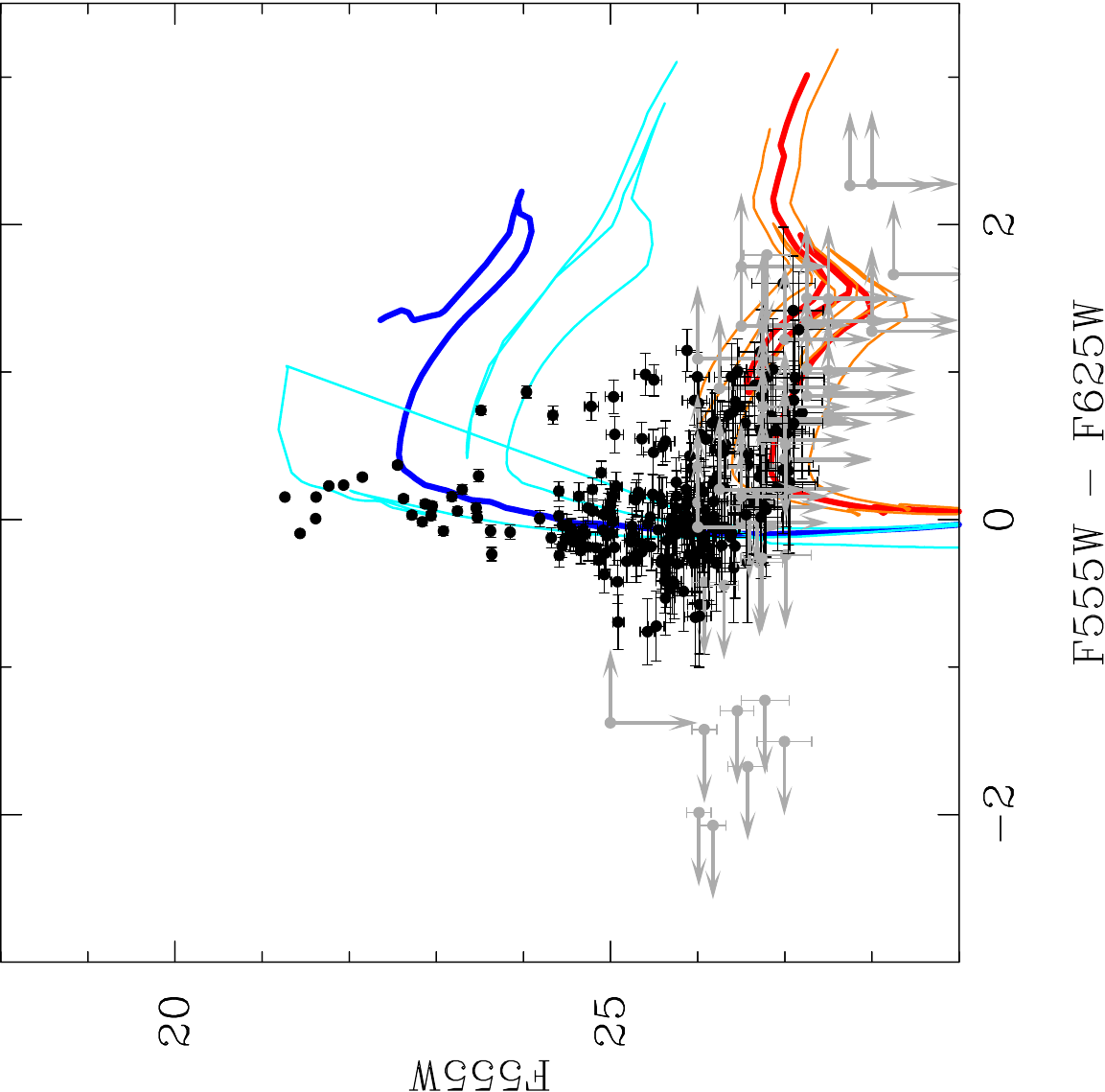}
\caption{Colour magnitude diagram for stars observed within 3 arcsec of the position of SN 2007gr in the HST WFC3/UVIS F336W, F555W and F625W bands.   The young and old stellar population solutions are indicated by the blue and red isochrones, respectively.  The central age $\tau$ for each solution is indicated by the heavy lines, while the $\pm \sigma_{\tau}$ isochrones are indicated by the lighter lines.  Points shown in light grey are those photometric datasets that include upper detection limits.}
\label{fig:cmd:wfc3}
\end{figure*}

\begin{figure*}
\includegraphics[height=8cm, angle=270]{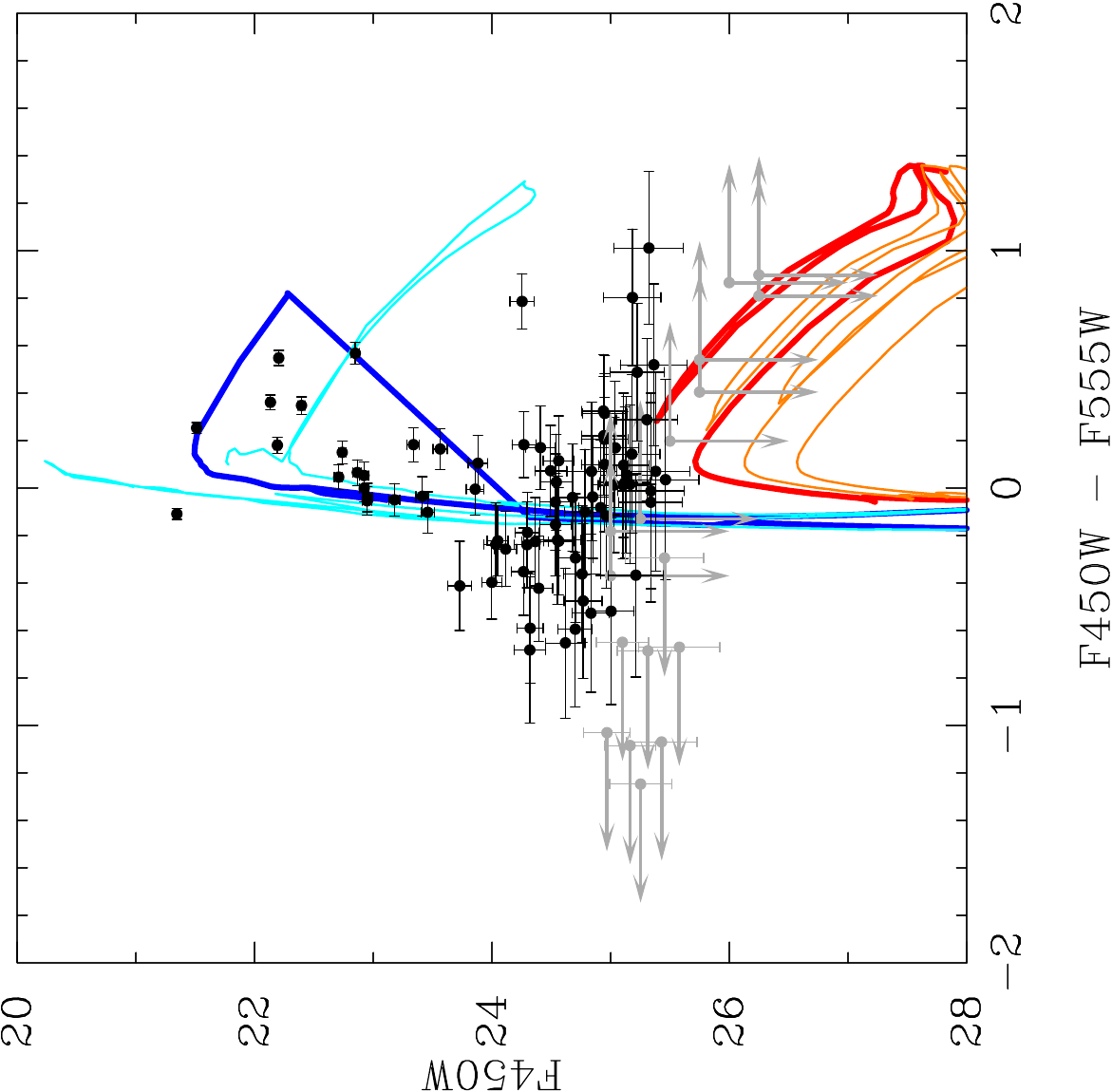}
\includegraphics[height=8cm, angle=270]{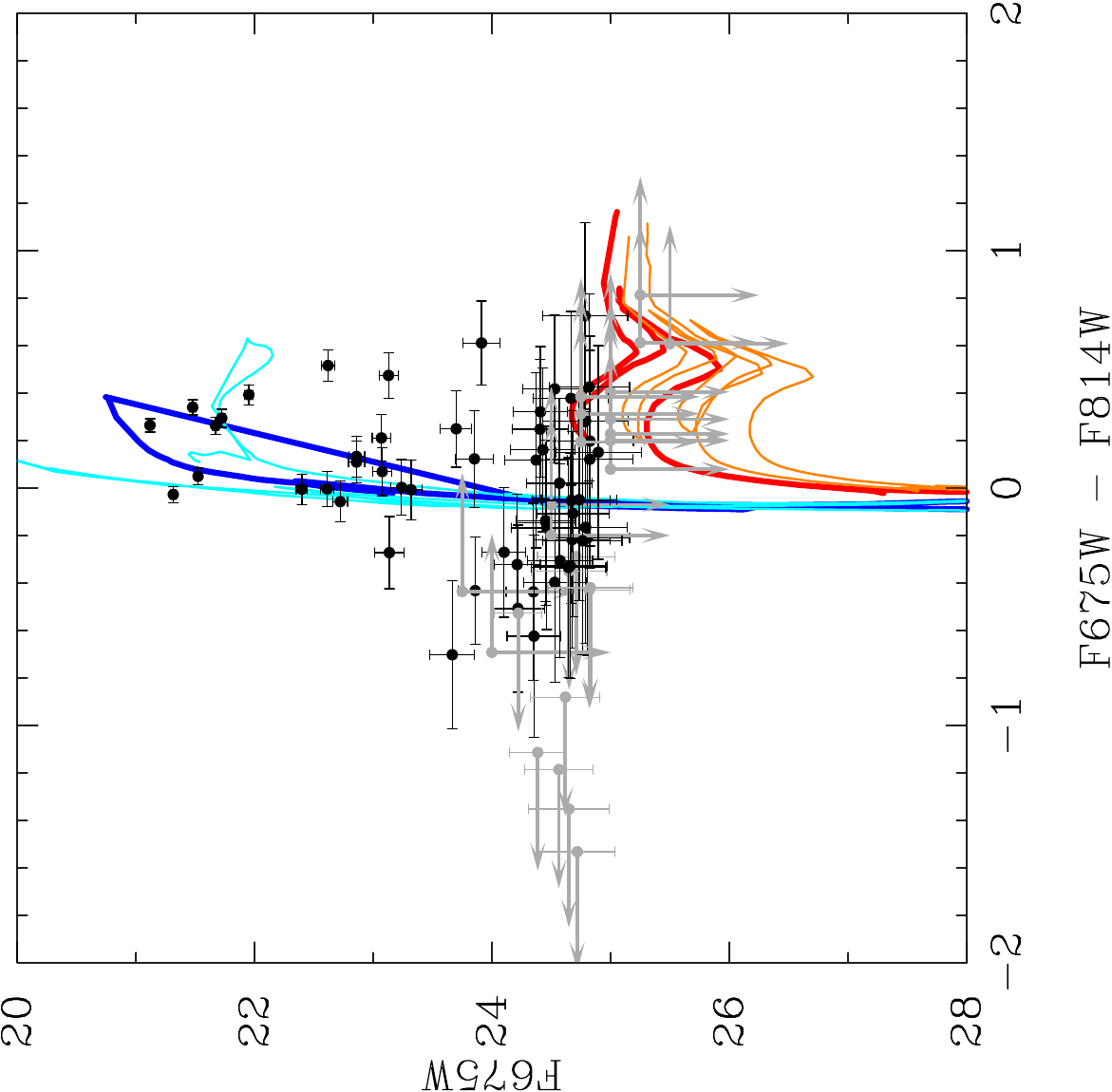}
\caption{The same as for Fig. \ref{fig:cmd:wfc3}, except showing colour magnitude diagrams for stars observed within 3 arcsec of the position of SN 2007gr in the HST WFPC2 F450W, F555W, F675W and F814W bands.}
\label{fig:cmd:wfpc2}
\end{figure*}

We applied this technique to the 284 and 99 detected stars in the late-time WFC3 and WFPC2 observations, respectively, within 3 arcsecs of the position of SN~2007gr.   Due to the differences in depth and spatial resolution between the WFPC2 and WFC3 observations, we kept the two datasets separate and conducted the analyses in parallel.  The positions of these stars in colour magnitude diagrams are shown on Figures \ref{fig:cmd:wfc3} and \ref{fig:cmd:wfpc2}.  The colour-magnitude diagrams for both sets of observations are dominated by a "blue plume" of bright stars and a fainter collection of stars consistent with the main sequence.

\begin{figure*}
\includegraphics[width=8cm, angle=270]{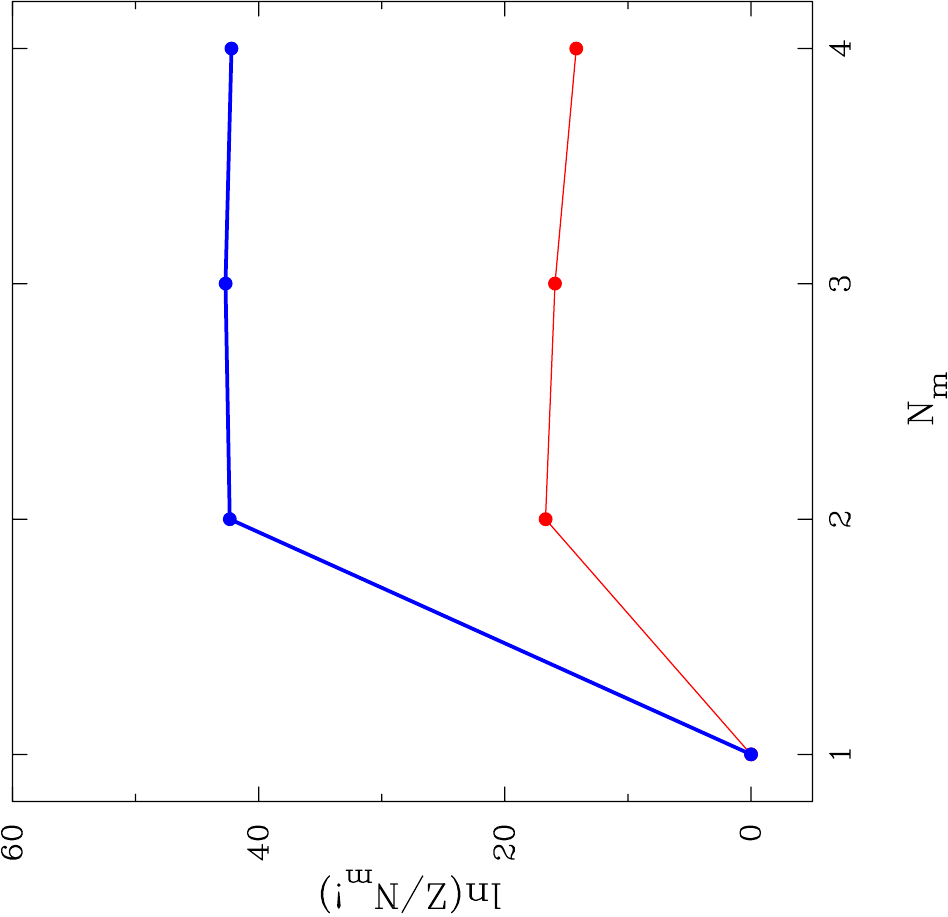}
\caption{Bayesian evidence values for stellar population fits for stars in the WFC3/UVIS (blue) and WFPC2 (red) observations for different numbers of mixture components ($N_{m}$).}
\label{fig:res:evidence}
\end{figure*}

\begin{figure*}
\includegraphics[width=8cm,angle=270]{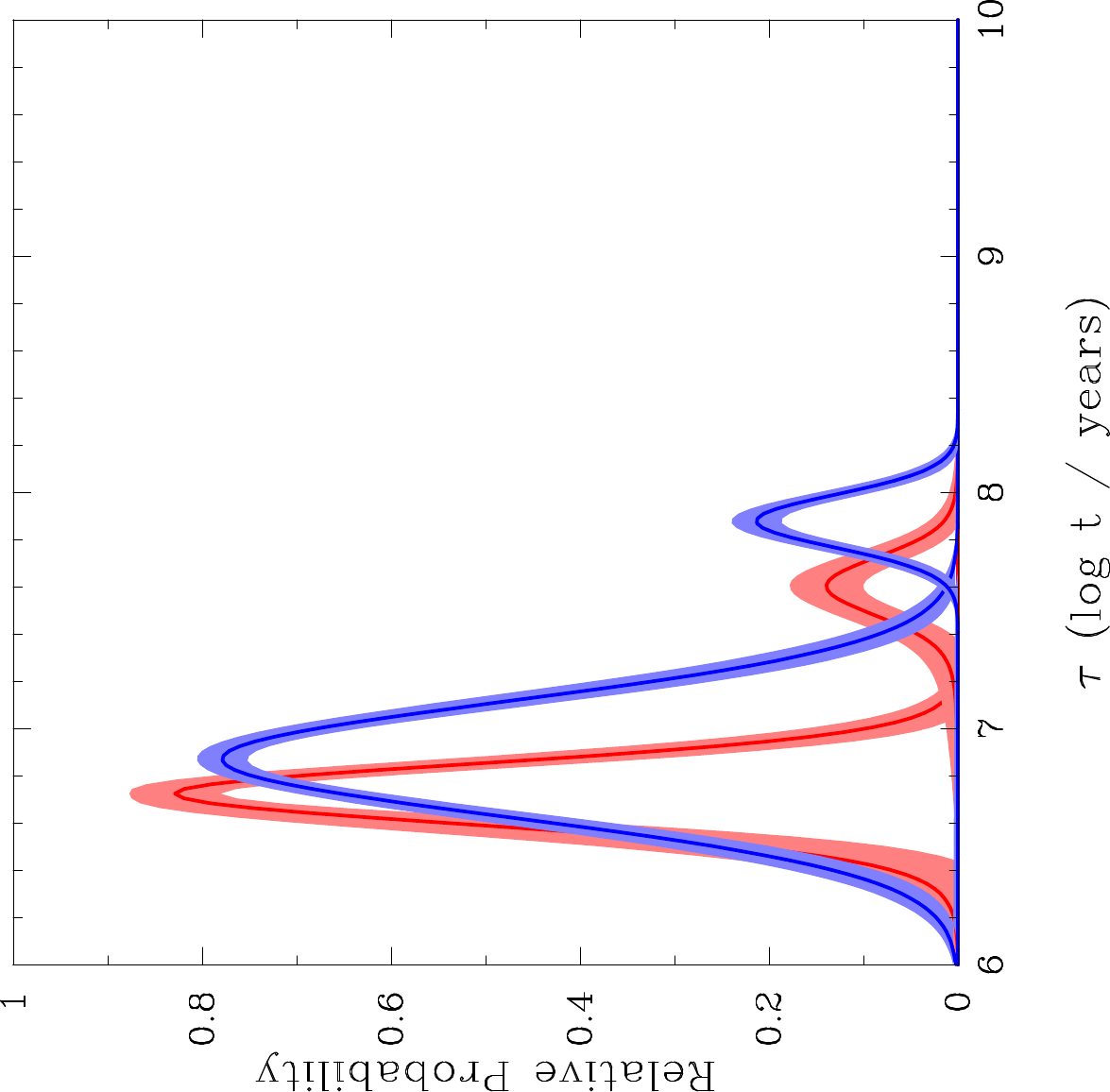}
\caption{The star formation history for the stellar population observed at the site of SN~2007gr for the optimal number of mixture components $N_{m} = 2$ for the WFC3/UVIS (blue) and WFPC2 (red) observations.  The heights of the peaks are scaled to correspond to the weighting of each star formation epoch.}
\label{fig:res:sfh}
\end{figure*}

\begin{figure*}
\includegraphics[height=8cm, angle=270]{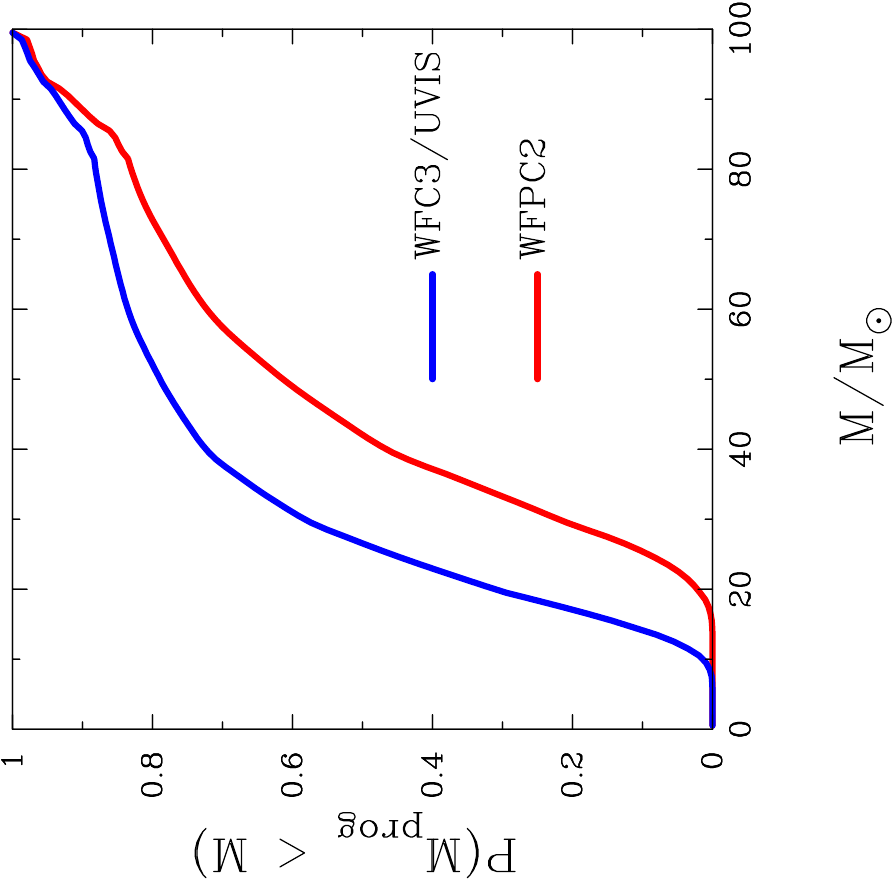}
\caption{Probability for the progenitor mass of SN 2007gr, given the parameters derived for the surrounding stellar population observed in HST WFC3 (blue) and WFPC2 (red) observations.}
\label{fig:res:mass}
\end{figure*}

We utilised Padova isochrones assuming a solar metallicity.  We used isochrones with ages in the range $6.00 \leq \tau \leq 10.00$ (at $\Delta \tau = 0.025$ intervals) for extinctions in the range $0.00 \leq A_{V} \leq 4.00$ (with a \citealt{ccm89} reddening law). The resulting values for the Bayesian evidence for  $N_{m} = 1 - 4$ mixture components are presented in Table \ref{tab:evid} and on Figure \ref{fig:res:evidence}, from which it is clear that the optimal value of mixture components to describe both datasets is $N_{m} = 2$.  The addition of further components does not significantly improve the quality of the fits to either the WFPC2 or WFC3 data.  The best fit parameters for the two populations are presented in Table \ref{tab:res:sfh} and the corresponding "effective" star formation histories are presented on Fig. \ref{fig:res:sfh}.   The position of the corresponding isochrones for the best fit solutions (and also the $\tau \pm 1\sigma_{\tau}$ solutions) are shown on Figures \ref{fig:cmd:wfc3} and \ref{fig:cmd:wfpc2}.  We find that that the WFPC2 observations suggest a slightly younger, less reddened population; however, this is likely due to the limited number of stars observed in the shallower WFPC2 observations.  For both sets of observations, we find evidence for an older stellar population solution that is composed of $\sim 1/5$ (by number) of the stars observed in the field.\\
Under the assumption that the progenitor of SN 2007gr arose from the observed stellar populations, given the corresponding dependence of a star's lifetime on its initial mass we can derive the probability for the initial mass of the progenitor.  On Figure \ref{fig:res:mass}, we show that WFPC2 observations, which imply younger, narrower epochs of star formation, support a higher mass progenitor than the WFC3 observations.  For the WFPC2 and WFC3 observations, the mass at which $P(M_{prog} < M) = 0.5$ occurs for $M = 41M_{\odot}$ and $27M_{\odot}$, respectively.  Both sets of observations, therefore, support a progenitor significantly more massive than those found for Type IIP or IIb SNe.


\begin{table*}
\caption{\label{tab:evid} Bayesian evidence values for fits, with varying number of mixture components ($N_{m}$),  to the observed WFC3 and WFPC2 data.}
\begin{tabular}{ccccc}
\hline\hline
                 &             & \multicolumn{3}{c}{$\ln Z / N_{m}!$$^{\dagger}$} \\
                 \\
\cline{3 - 3}\cline{5 - 5}
$N_{m}$   & $n_{par}$  & WFC3/UVIS & & WFPC2 \\
1               &3           & $0.00\pm0.15$ & & $0.00 \pm 0.13$\\
2              & 7          & $42.34\pm0.18$ & & $ 16.70 \pm 0.18$  \\
3               &10        & $42.70\pm0.18$ & & $15.92 \pm 0.18$   \\
4               & 13      & $42.20\pm0.18$ & & $14.19 \pm 0.19$   \\
\hline\hline
\end{tabular}\\
$^{\dagger}$ Normalised with respect to $N_{m} = 1$.
\end{table*}

\begin{table*}
\caption{\label{tab:res:sfh}Recovered star formation histories for the stellar population observed in WFC3/UVIS and WFPC2 observations of the site of SN~2007gr.}
\begin{tabular}{cccccccccc}
\hline\hline
 & & & \multicolumn{3}{c}{Young} & & \multicolumn{3}{c}{Old} \\
 \cline{4 - 6}\cline{8 - 10}\\
 & $A_{V}$ & &$\tau_{1}$ & $\sigma_{1}$ & $w_{1}$ & & $\tau_{2}$ & $\sigma_{2}$ & $w_{2}$ \\
\hline
WFC3/UVIS &  $0.53 \pm 0.00$ & & $6.87 \pm 0.02$ & $0.25 \pm 0.02$ & $0.78 \pm 0.03$ & & $7.88 \pm 0.02$ & $0.11 \pm 0.01$ & $0.22 \pm 0.02$\\
WFPC2 & $0.44 \pm 0.01$ & & $6.71 \pm 0.03$ & $0.14 \pm 0.03$ & $0.85 \pm 0.03$ & & $7.60 \pm 0.03$ & $0.09 \pm 0.03 $ & $0.15 \pm 0.02$\\
\hline\hline
\end{tabular}
\end{table*}

\section{Discussion and Conclusions}
\label{sec:disc}

Despite the lack of a detection of a progenitor in pre-explosion images, the nature of the progenitor of SN 2007gr is particularly intriguing given the classification of the SN as a Type Ic SN and the location of the SN at almost the centre of a dense young, massive star association and in close proximity to a star as massive as $40M_{\odot}$.  From the WFC3 and WFPC2 observations there  is immediate support for the hypothesis that SN 2007gr resulted from a Wolf-Rayet progenitor that was an initially massive star.

SN 2007gr is an important Type Ic SN, principally because of the proximity of its host galaxy which resulted in a detailed photometric and spectroscopic monitoring campaign \citep{2008ApJ...673L.155V, 2009A&A...508..371H}. Studies by \citet{2010Natur.463..516P} and \citet{2011ApJ...735....3X} reported that SN 2007gr also hosted a relativistic outflow, that might imply SN 2007gr had a connection with the Gamma Ray Burst phenomenon.  Using further X-ray and radio observations of SN 2007gr, \citet{2010ApJ...725..922S} robustly disagree with the previous conclusions and find SN 2007gr to only be expanding at trans-relativistic ($\approx 0.2 $c) velocities.  SN 2007gr is extraordinary for its very prompt  radio peak in luminosity which, in comparison with the compilation of  observations of Type II (including IIP, IIn and IIL) and Type  Ib/c  SNe compiled by \citet{2015arXiv150407988K}, implies high blast wave speeds that are only compatible with compact progenitors such as WR stars.

Based on previous HST observations of the site of SN 2007gr, \citet{2008ApJ...672L..99C}  and \citet{2014ApJ...790..120C} have presented mass estimates based on ages determined from the SED of Source A under the assumption that it was a cluster. Based on the high-resolution HST images presented here, we find Source A is consistent with a point stellar-like source, and not a cluster; hence, the previously derived masses for the progenitor are based on an assumption that we propose to be flawed.   From a census of the ejecta, conducted using nebular phase spectroscopy of SN 2007gr, \citet{2010MNRAS.408...87M} calculated that the supernova resulted from the explosion of a relatively low mass star with $M_{ZAMS} \approx 15M_{\odot}$.  This is at the lower limit for the masses we estimate for the progenitor from the age of the surrounding stellar population.  Given the radically different nature of the two approaches, the source of the discrepancy between our mass estimate for the progenitor and the estimate presented by \citet{2010MNRAS.408...87M} is unclear.

\begin{figure*}
\includegraphics[height=6.5cm, angle=270]{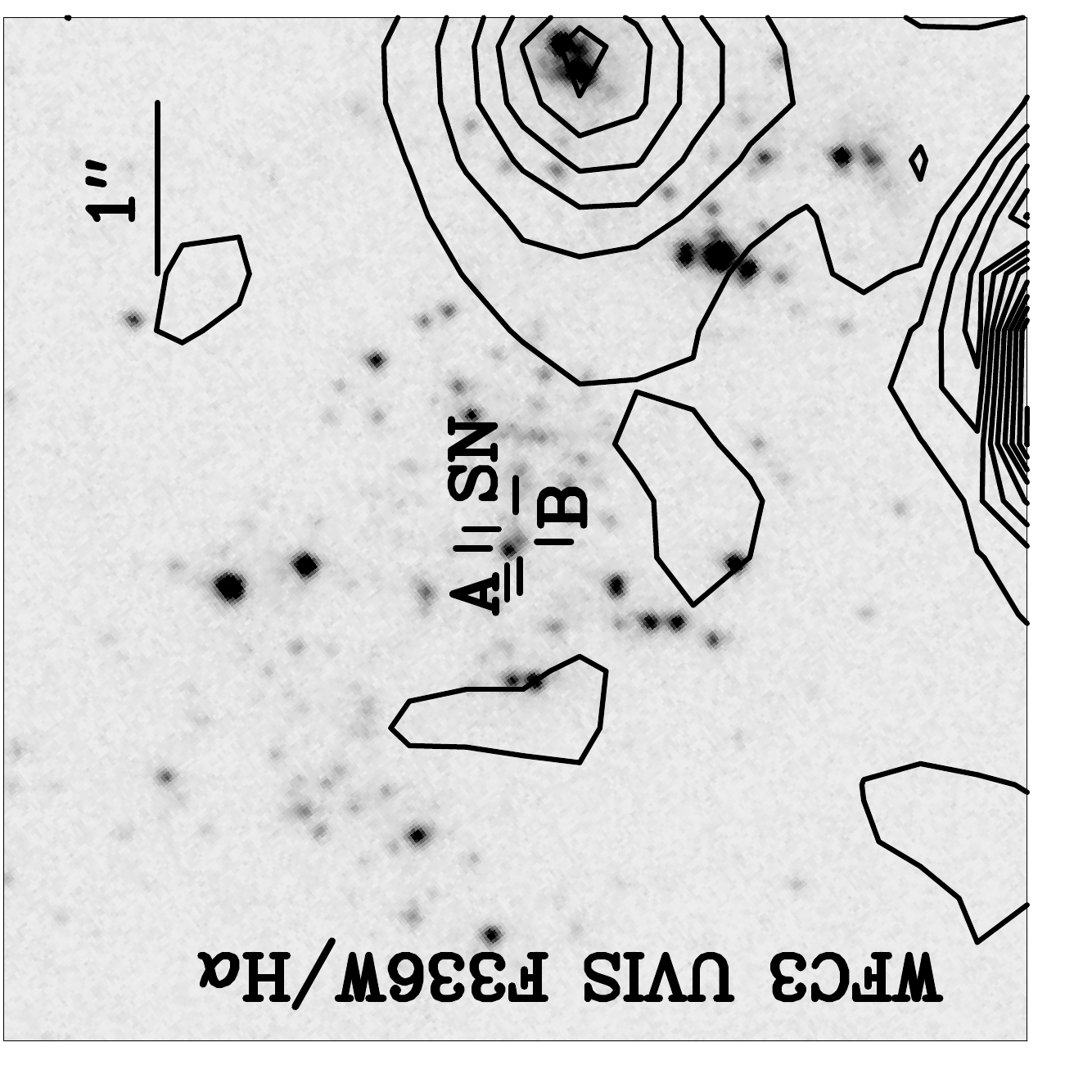}
\caption{Late-time WFC3/UVIS observation of the site of SN 2007gr with contours overlaid corresponding to a continuum-subtracted INT WFC narrow band  $H\alpha$ image.}
\label{fig:disc:halpha}
\end{figure*}

 \citet{2013AJ....146...30K} used integral field spectroscopy of knots of $H\alpha$ emission surrounding the site of SN 2007gr to estimate an age for the progenitor in the range of $\tau = 6.3-7.8$ using the equivalent width of $H\alpha$ as a proxy for the age. The reported age range is consistent with the age for the stellar population found here.  \citeauthor{2013AJ....146...30K} note, however, that given the limitations of the seeing conditions of the observations, individual sources in the vicinity of the SN were not resolved.  Previously \citet{2008ApJ...672L..99C} discussed a continuum subtracted $H\alpha$ image of the site of SN 2007gr acquired on 2005 January 13 with the Wide Field Camera (WFC) of the Isaac Newton Telescope (INT).   On Figure \ref{fig:disc:halpha},  we show contours corresponding to $H\alpha$ emission over the late-time WFC3/UVIS F336W observation.  Despite the significant difference in spatial resolution between the two images, it is clear that the site of SN 2007gr and the centre of the massive star association are not associated with significant $H\alpha$ emission.  This has important consequences for the use of the $H\alpha$ equivalent width as an age proxy, since the SN position may itself not be associated with any $H\alpha$ emission.  The use of pixel statistics to assess the proximity of SN positions with tracers of star formation (e.g. $H\alpha$ emission; for a review see \citealt{2015PASA...32...19A}) may also be undermined by the use of low resolution ground based imaging which, as in the case of SN 2007gr, might correlate $H\alpha$ emission with a SN, where there is actually no association; further reinforcing the conclusions of \citet{2013MNRAS.428.1927C}.  This highlights the fact that weak association with $H\alpha$ emission does not immediately imply a low mass progenitor.\\

Here, we have used observations of the surrounding stellar population covering a large wavelength range from the ultraviolet to the near-infrared.  The observations, as presented in Fig. \ref{fig:cmd:wfc3} and \ref{fig:cmd:wfpc2}, probe a large section of the main sequence of massive stars, as well as a small number of stars that have evolved off the main sequence.  The only requirement of this analysis is that the unseen progenitor is drawn from the same underlying distributions as the observed population of stars.    Previously, studies of the ages of resolved stellar populations associated with the sites of CCSNe have concentrated on the nearby population within a radius of $\mathrm{\sim 50\,pc}$, under the assumption that stars in such proximity will be effectively co-eval with the progenitor, while at the same time limiting contamination from the background population \citep{2009ApJ...703..300G}.  We have not placed a strict radius constraint on the size of the region associated with SN 2007gr, except by eye constraining the size of the massive star association to $\sim 300\,\mathrm{pc}$.  The Bayesian analysis presented here permits us to include as many additional populations as required to fit the observed data.  In addition, a background stellar population could also be treated as an additional mixture component, described by stars away from the central region of interest.

We have approximately estimated the spatial extent of the massive stellar population with which SN 2007gr is associated to have a diameter of $\sim 300\mathrm{pc}$.  The ages we have derived from the WFPC2 and WFC3 observations are significantly shorter than the expected sound crossing for a single cloud of this size \citep[$\sim 1 \,\mathrm{Myr\,pc^{-1}}$][]{2000ApJ...530..277E}.    There are, however, massive star complexes that cover large spatial scales in the Galaxy, that have small age spreads that are also incompatible with the expected sound crossing times.  By comparison \citet[][and references therein]{2002AJ....124..404P} studied the OB association U Sco, which has an approximate spatial extent of $\sim \mathrm{50 - 70\,pc}$, and found the stellar population could be described by a single age of 5 Myr, without any significant age spread, implying a single burst of star formation.  \citeauthor{2002AJ....124..404P} suggested that star formation was triggered by a SN shock wave.   Although U Sco has smaller dimensions than the massive star association in NGC 1058, it does illustrate that such large structures with small age spreads can exist.   Conversely, if the constituent stars of the association were formed in a single location, velocity dispersions of only $10 - 20\,\mathrm{km\,s^{-1}}$ would be required to reproduce the observed spatial distribution of stars in the lifetime of the young age solution.   It is also worth contrasting the dense environment in which the, apparently, massive progenitor of SN~2007gr resides, compared to the relatively sparse neighbourhood of SN 2002ap, for which deep pre-explosion images ruled out all single massive progenitor scenarios \citep{2007MNRAS.381..835C,2013MNRAS.436..774E}.

An important factor not included in our Bayesian treatment is the possibility of differential extinction.  We have assumed that the stars that make up the two observed populations  share the same, single valued reddening.  We note that, similarly to the differences in age inferred from the WFC3 and WFPC2 observations, the difference in wavelength coverage may be behind the difference in extinction; with observations in the UV more heavily affected by reddening, than the corresponding redder WFPC2 observations.  If  the extinction for each star arises from a distribution of extinctions, rather than a single value, this could also be implemented using an hierarchical scheme.  Care, however, would need to be taken to avoid any biases introduced due to the requirement that $A_{V} \geq 0$.  It is expected that the inclusion of differential reddening would make the age distributions narrower; as a single isochrone could accommodate a larger range of colours, than under the assumption of a single value for the extinction.

Such analysis of the host environments of SNe, as presented here, has been previously employed by \citet{2009ApJ...703..300G}, \citet{2011ApJ...742L...4M},  and \citet{2014ApJ...795..170J} for estimating the progenitor masses for CCSNe.  \citet{2014ApJ...791..105W} found that for the Type IIP SNe in their sample they were able to achieve similar estimates for the median mass as those derived directly for the progenitor in pre-explosion images; as was also achieved by \citet{2011ApJ...742L...4M} from their analysis of the population surrounding the Type IIb SN 2011dh.  Although, as evidenced by Figure \ref{fig:res:mass}, there are large uncertainties associated with this approach, the case of SN 2007gr does highlight that the lack of a detection of a progenitor in pre-explosion images does not imply a low mass.  \citet{2015PASA...32...16S} argues that the lack of detection of progenitors of high mass stars ($\gtrsim 18M_{\odot}$) suggests that these stars may not be exploding at all as SNe, but rather collapsing to form black holes.  Caution is required in drawing such a conclusion as there are a number of major issues concerning the analysis of the pre-explosion detections of the progenitors of CCSNe that have yet to be resolved.  A fundamental problem in the consideration of the ensemble of progenitor detections and non-detections in a statistical context is that it is unclear how the, generally poor, quality of available pre-explosion observations effectively censors the underlying distribution of SN progenitors.  In the case of the pre-explosion observations of SN 2007gr, consisting of shallow WFPC2 F450W and F814W images, it was not surprising that a possible Wolf-Rayet progenitor might not be detected \citep{2008ApJ...672L..99C}.  We used the SEDs of \citet{2012A&A...540A.144S}\footnote{http://www.astro.physik.uni-potsdam.de/$\sim$wrh/PoWR/powrgrid1.html} (over a temperature range of $45\,000 \leq T \leq 200\,000\,\mathrm{K}$ and transformed radius of $-0.5\leq  \log (R_{t}/R_{\odot}) \leq 1.6$) to probe the sensitivity of the pre-explosion observations (see Section \ref{sec:res:location}) to a WC progenitor.  Using the extinction estimate for the host stellar population determined here,  the minimum luminosity that might yield a detectable progenitor ($p(\mathrm{detect}) = 0.5$) in the pre-explosion observations is $\log (L/L_{\odot}) \sim 6.5$.  This limit is significantly higher than the luminosities predicted for stars that will end their lives as WR stars \citep[$\log (L/L_{\odot}) < 6.0$;][]{eld04,2012A&A...544L..11Y} and the luminosities derived by \citeauthor{2012A&A...540A.144S} for a sample of Galactic WC stars.  Based on stellar populations associated with SN remnants in M31 and M33, \citet{2014ApJ...795..170J} conclude that the maximum mass for a star to explode may be as high as $\sim 35-45M_{\odot}$.   In the absence of pre-explosion data of sufficient quality to detect the progenitors of certain types of SNe, in particular those from higher mass stars, this analysis could play an important role in establishing the fate of those stars that do not lead to progenitors as easy to detect as RSGs.

It is clear from analyses such as this that contextual information about the locations of SN explosions can be just as important as direct detection of progenitors.  Indeed, if a progenitor is detected a significant amount of information might be missed by ignoring the stars nearest to it.  In studying population of stars associated with a SN, we may potentially directly probe the physical conditions in which the progenitor formed and in which its entire evolution took place; to which some observational proxies (such as $H\alpha$ emission equivalent width) might be insensitive in individual cases, only providing insight for a large statistical sample.

\section*{Acknowledgements}
The research of JRM is supported through Royal Society University Research Fellowship.  The authors thank Daniel Mortlock for reviewing a draft of the paper and to the referee Nancy Elias-Rosa.  The authors are grateful to Jason Kalirai, Paul Crowther and Simon Goodwin for their useful discussions during the preparation of this manuscript.  
\bibliographystyle{apj}

\begin{thebibliography}{66}
\expandafter\ifx\csname natexlab\endcsname\relax\def\natexlab#1{#1}\fi

\bibitem[{{Anderson} {et~al.}(2015){Anderson}, {James}, {Habergham}, {Galbany},
  \& {Kuncarayakti}}]{2015PASA...32...19A}
{Anderson}, J.~P., {James}, P.~A., {Habergham}, S.~M., {Galbany}, L., \&
  {Kuncarayakti}, H. 2015, \pasa, 32, 19

\bibitem[{{Asplund} {et~al.}(2004){Asplund}, {Grevesse}, {Sauval}, {Allende
  Prieto}, \& {Kiselman}}]{2004A&A...417..751A}
{Asplund}, M., {Grevesse}, N., {Sauval}, A.~J., {Allende Prieto}, C., \&
  {Kiselman}, D. 2004, \aap, 417, 751

\bibitem[{{Bastian} {et~al.}(2005){Bastian}, {Gieles}, {Lamers}, {Scheepmaker},
  \& {de Grijs}}]{2005A&A...431..905B}
{Bastian}, N., {Gieles}, M., {Lamers}, H.~J.~G.~L.~M., {Scheepmaker}, R.~A., \&
  {de Grijs}, R. 2005, \aap, 431, 905

\bibitem[{{Bersten} {et~al.}(2014){Bersten}, {Benvenuto}, {Folatelli},
  {Nomoto}, {Kuncarayakti}, {Srivastav}, {Anupama}, {Quimby}, \&
  {Sahu}}]{2014AJ....148...68B}
{Bersten}, M.~C., {Benvenuto}, O.~G., {Folatelli}, G., {Nomoto}, K.,
  {Kuncarayakti}, H., {Srivastav}, S., {Anupama}, G.~C., {Quimby}, R., \&
  {Sahu}, D.~K. 2014, \aj, 148, 68

\bibitem[{{Bressan} {et~al.}(2013){Bressan}, {Marigo}, {Girardi}, {Nanni}, \&
  {Rubele}}]{2013EPJWC..4303001B}
{Bressan}, A., {Marigo}, P., {Girardi}, L., {Nanni}, A., \& {Rubele}, S. 2013,
  in European Physical Journal Web of Conferences, Vol.~43, European Physical
  Journal Web of Conferences, 3001

\bibitem[{{Bressan} {et~al.}(2012){Bressan}, {Marigo}, {Girardi}, {Salasnich},
  {Dal Cero}, {Rubele}, \& {Nanni}}]{2012MNRAS.427..127B}
{Bressan}, A., {Marigo}, P., {Girardi}, L., {Salasnich}, B., {Dal Cero}, C.,
  {Rubele}, S., \& {Nanni}, A. 2012, \mnras, 427, 127

\bibitem[{{Cao} {et~al.}(2013){Cao}, {Kasliwal}, {Arcavi}, {Horesh}, {Hancock},
  {Valenti}, {Cenko}, {Kulkarni}, {Gal-Yam}, {Gorbikov}, {Ofek}, {Sand},
  {Yaron}, {Graham}, {Silverman}, {Wheeler}, {Marion}, {Walker}, {Mazzali},
  {Howell}, {Li}, {Kong}, {Bloom}, {Nugent}, {Surace}, {Masci}, {Carpenter},
  {Degenaar}, \& {Gelino}}]{2013ApJ...775L...7C}
{Cao}, Y., {Kasliwal}, M.~M., {Arcavi}, I., {Horesh}, A., {Hancock}, P.,
  {Valenti}, S., {Cenko}, S.~B., {Kulkarni}, S.~R., {Gal-Yam}, A., {Gorbikov},
  E., {Ofek}, E.~O., {Sand}, D., {Yaron}, O., {Graham}, M., {Silverman}, J.~M.,
  {Wheeler}, J.~C., {Marion}, G.~H., {Walker}, E.~S., {Mazzali}, P., {Howell},
  D.~A., {Li}, K.~L., {Kong}, A.~K.~H., {Bloom}, J.~S., {Nugent}, P.~E.,
  {Surace}, J., {Masci}, F., {Carpenter}, J., {Degenaar}, N., \& {Gelino},
  C.~R. 2013, \apjl, 775, L7

\bibitem[{{Cardelli} {et~al.}(1989){Cardelli}, {Clayton}, \& {Mathis}}]{ccm89}
{Cardelli}, J.~A., {Clayton}, G.~C., \& {Mathis}, J.~S. 1989, \apj, 345, 245

\bibitem[{{Castelli} \& {Kurucz}(2004)}]{2004astro.ph..5087C}
{Castelli}, F., \& {Kurucz}, R.~L. 2004, ArXiv Astrophysics e-prints 0405087

\bibitem[{{Chen} {et~al.}(2014){Chen}, {Wang}, {Ganeshalingam}, {Silverman},
  {Filippenko}, {Li}, {Chornock}, {Li}, \& {Steele}}]{2014ApJ...790..120C}
{Chen}, J., {Wang}, X., {Ganeshalingam}, M., {Silverman}, J.~M., {Filippenko},
  A.~V., {Li}, W., {Chornock}, R., {Li}, J., \& {Steele}, T. 2014, \apj, 790,
  120

\bibitem[{{Crockett} {et~al.}(2008){Crockett}, {Maund}, {Smartt}, {Mattila},
  {Pastorello}, {Smoker}, {Stephens}, {Fynbo}, {Eldridge}, {Danziger}, \&
  {Benn}}]{2008ApJ...672L..99C}
{Crockett}, R.~M., {Maund}, J.~R., {Smartt}, S.~J., {Mattila}, S.,
  {Pastorello}, A., {Smoker}, J., {Stephens}, A.~W., {Fynbo}, J., {Eldridge},
  J.~J., {Danziger}, I.~J., \& {Benn}, C.~R. 2008, \apjl, 672, L99

\bibitem[{{Crockett} {et~al.}(2007){Crockett}, {Smartt}, {Eldridge}, {Mattila},
  {Young}, {Pastorello}, {Maund}, {Benn}, \& {Skillen}}]{2007MNRAS.381..835C}
{Crockett}, R.~M., {Smartt}, S.~J., {Eldridge}, J.~J., {Mattila}, S., {Young},
  D.~R., {Pastorello}, A., {Maund}, J.~R., {Benn}, C.~R., \& {Skillen}, I.
  2007, \mnras, 381, 835

\bibitem[{{Crowther}(2013)}]{2013MNRAS.428.1927C}
{Crowther}, P.~A. 2013, \mnras, 428, 1927

\bibitem[{{Dolphin}(2000)}]{dolphhstphot}
{Dolphin}, A.~E. 2000, \pasp, 112, 1383

\bibitem[{{Eldridge} {et~al.}(2015){Eldridge}, {Fraser}, {Maund}, \&
  {Smartt}}]{2015MNRAS.446.2689E}
{Eldridge}, J.~J., {Fraser}, M., {Maund}, J.~R., \& {Smartt}, S.~J. 2015,
  \mnras, 446, 2689

\bibitem[{{Eldridge} {et~al.}(2013){Eldridge}, {Fraser}, {Smartt}, {Maund}, \&
  {Crockett}}]{2013MNRAS.436..774E}
{Eldridge}, J.~J., {Fraser}, M., {Smartt}, S.~J., {Maund}, J.~R., \&
  {Crockett}, R.~M. 2013, \mnras, 436, 774

\bibitem[{{Eldridge} \& {Tout}(2004)}]{eld04}
{Eldridge}, J.~J., \& {Tout}, C.~A. 2004, \mnras, 353, 87

\bibitem[{{Elmegreen}(2000)}]{2000ApJ...530..277E}
{Elmegreen}, B.~G. 2000, \apj, 530, 277

\bibitem[{{Ferguson} {et~al.}(1998){Ferguson}, {Gallagher}, \&
  {Wyse}}]{1998AJ....116..673F}
{Ferguson}, A.~M.~N., {Gallagher}, J.~S., \& {Wyse}, R.~F.~G. 1998, \aj, 116,
  673

\bibitem[{{Feroz} \& {Hobson}(2008)}]{2008MNRAS.384..449F}
{Feroz}, F., \& {Hobson}, M.~P. 2008, \mnras, 384, 449

\bibitem[{{Feroz} {et~al.}(2013){Feroz}, {Hobson}, {Cameron}, \&
  {Pettitt}}]{2013arXiv1306.2144F}
{Feroz}, F., {Hobson}, M.~P., {Cameron}, E., \& {Pettitt}, A.~N. 2013, ArXiv
  e-prints: 1306.2144

\bibitem[{{Foley} {et~al.}(2011){Foley}, {Berger}, {Fox}, {Levesque},
  {Challis}, {Ivans}, {Rhoads}, \& {Soderberg}}]{2011ApJ...732...32F}
{Foley}, R.~J., {Berger}, E., {Fox}, O., {Levesque}, E.~M., {Challis}, P.~J.,
  {Ivans}, I.~I., {Rhoads}, J.~E., \& {Soderberg}, A.~M. 2011, \apj, 732, 32

\bibitem[{{Fremling} {et~al.}(2014){Fremling}, {Sollerman}, {Taddia}, {Ergon},
  {Valenti}, {Arcavi}, {Ben-Ami}, {Cao}, {Cenko}, {Filippenko}, {Gal-Yam}, \&
  {Howell}}]{2014A&A...565A.114F}
{Fremling}, C., {Sollerman}, J., {Taddia}, F., {Ergon}, M., {Valenti}, S.,
  {Arcavi}, I., {Ben-Ami}, S., {Cao}, Y., {Cenko}, S.~B., {Filippenko}, A.~V.,
  {Gal-Yam}, A., \& {Howell}, D.~A. 2014, \aap, 565, A114

\bibitem[{{Gal-Yam} \& {Leonard}(2009)}]{galyam05gl}
{Gal-Yam}, A., \& {Leonard}, D.~C. 2009, \nat, 458, 865

\bibitem[{{Gal-Yam} {et~al.}(2007){Gal-Yam}, {Leonard}, {Fox}, {Cenko},
  {Soderberg}, {Moon}, {Sand}, {Li}, {Filippenko}, {Aldering}, \&
  {Copin}}]{2007ApJ...656..372G}
{Gal-Yam}, A., {Leonard}, D.~C., {Fox}, D.~B., {Cenko}, S.~B., {Soderberg},
  A.~M., {Moon}, D.-S., {Sand}, D.~J., {Li}, W., {Filippenko}, A.~V.,
  {Aldering}, G., \& {Copin}, Y. 2007, \apj, 656, 372

\bibitem[{{Girardi} {et~al.}(2002){Girardi}, {Bertelli}, {Bressan}, {Chiosi},
  {Groenewegen}, {Marigo}, {Salasnich}, \& {Weiss}}]{2002A&A...391..195G}
{Girardi}, L., {Bertelli}, G., {Bressan}, A., {Chiosi}, C., {Groenewegen},
  M.~A.~T., {Marigo}, P., {Salasnich}, B., \& {Weiss}, A. 2002, \aap, 391, 195

\bibitem[{{Gogarten} {et~al.}(2009){Gogarten}, {Dalcanton}, {Murphy},
  {Williams}, {Gilbert}, \& {Dolphin}}]{2009ApJ...703..300G}
{Gogarten}, S.~M., {Dalcanton}, J.~J., {Murphy}, J.~W., {Williams}, B.~F.,
  {Gilbert}, K., \& {Dolphin}, A. 2009, \apj, 703, 300

\bibitem[{{Groh} {et~al.}(2013{\natexlab{a}}){Groh}, {Georgy}, \&
  {Ekstr{\"o}m}}]{2013A&A...558L...1G}
{Groh}, J.~H., {Georgy}, C., \& {Ekstr{\"o}m}, S. 2013{\natexlab{a}}, \aap,
  558, L1

\bibitem[{{Groh} {et~al.}(2013{\natexlab{b}}){Groh}, {Meynet}, {Georgy}, \&
  {Ekstr{\"o}m}}]{2013A&A...558A.131G}
{Groh}, J.~H., {Meynet}, G., {Georgy}, C., \& {Ekstr{\"o}m}, S.
  2013{\natexlab{b}}, \aap, 558, A131

\bibitem[{{Hunter} {et~al.}(2009){Hunter}, {Valenti}, {Kotak}, {Meikle},
  {Taubenberger}, {Pastorello}, {Benetti}, {Stanishev}, {Smartt}, {Trundle},
  {Arkharov}, {Bufano}, {Cappellaro}, {Di Carlo}, {Dolci}, {Elias-Rosa},
  {Frandsen}, {Fynbo}, {Hopp}, {Larionov}, {Laursen}, {Mazzali}, {Navasardyan},
  {Ries}, {Riffeser}, {Rizzi}, {Tsvetkov}, {Turatto}, \&
  {Wilke}}]{2009A&A...508..371H}
{Hunter}, D.~J., {Valenti}, S., {Kotak}, R., {Meikle}, W.~P.~S.,
  {Taubenberger}, S., {Pastorello}, A., {Benetti}, S., {Stanishev}, V.,
  {Smartt}, S.~J., {Trundle}, C., {Arkharov}, A.~A., {Bufano}, F.,
  {Cappellaro}, E., {Di Carlo}, E., {Dolci}, M., {Elias-Rosa}, N., {Frandsen},
  S., {Fynbo}, J.~U., {Hopp}, U., {Larionov}, V.~M., {Laursen}, P., {Mazzali},
  P., {Navasardyan}, H., {Ries}, C., {Riffeser}, A., {Rizzi}, L., {Tsvetkov},
  D.~Y., {Turatto}, M., \& {Wilke}, S. 2009, \aap, 508, 371

\bibitem[{{Izzard} {et~al.}(2004){Izzard}, {Ramirez-Ruiz}, \& {Tout}}]{izzgrb}
{Izzard}, R.~G., {Ramirez-Ruiz}, E., \& {Tout}, C.~A. 2004, \mnras, 348, 1215

\bibitem[{Jeffreys(1961)}]{Jeffreys61}
Jeffreys, H. 1961, Theory of Probability, 3rd edn. (Oxford, England: Oxford)

\bibitem[{{Jennings} {et~al.}(2014){Jennings}, {Williams}, {Murphy},
  {Dalcanton}, {Gilbert}, {Dolphin}, {Weisz}, \&
  {Fouesneau}}]{2014ApJ...795..170J}
{Jennings}, Z.~G., {Williams}, B.~F., {Murphy}, J.~W., {Dalcanton}, J.~J.,
  {Gilbert}, K.~M., {Dolphin}, A.~E., {Weisz}, D.~R., \& {Fouesneau}, M. 2014,
  \apj, 795, 170

\bibitem[{{J{\o}rgensen} \& {Lindegren}(2005)}]{2005A&A...436..127J}
{J{\o}rgensen}, B.~R., \& {Lindegren}, L. 2005, \aap, 436, 127

\bibitem[{{Kamble} {et~al.}(2015){Kamble}, {Margutti}, {Soderberg},
  {Chakraborti}, {Fransson}, {Chevalier}, {Powell}, {Milisavljevic}, {Parrent},
  \& {Bietenholz}}]{2015arXiv150407988K}
{Kamble}, A., {Margutti}, R., {Soderberg}, A.~M., {Chakraborti}, S.,
  {Fransson}, C., {Chevalier}, R., {Powell}, D., {Milisavljevic}, D.,
  {Parrent}, J., \& {Bietenholz}, M. 2015, ArXiv e-prints: 150407988

\bibitem[{{Kochanek} {et~al.}(2008){Kochanek}, {Beacom}, {Kistler}, {Prieto},
  {Stanek}, {Thompson}, \& {Y{\"u}ksel}}]{2008ApJ...684.1336K}
{Kochanek}, C.~S., {Beacom}, J.~F., {Kistler}, M.~D., {Prieto}, J.~L.,
  {Stanek}, K.~Z., {Thompson}, T.~A., \& {Y{\"u}ksel}, H. 2008, \apj, 684, 1336

\bibitem[{{Kuncarayakti} {et~al.}(2013){Kuncarayakti}, {Doi}, {Aldering},
  {Arimoto}, {Maeda}, {Morokuma}, {Pereira}, {Usuda}, \&
  {Hashiba}}]{2013AJ....146...30K}
{Kuncarayakti}, H., {Doi}, M., {Aldering}, G., {Arimoto}, N., {Maeda}, K.,
  {Morokuma}, T., {Pereira}, R., {Usuda}, T., \& {Hashiba}, Y. 2013, \aj, 146,
  30

\bibitem[{{Larsen}(1999)}]{1999A&AS..139..393L}
{Larsen}, S.~S. 1999, \aaps, 139, 393

\bibitem[{{Larsen}(2004)}]{2004A&A...416..537L}
---. 2004, \aap, 416, 537

\bibitem[{{Leitherer} {et~al.}(1999){Leitherer}, {Schaerer}, {Goldader},
  {Gonz{\'a}lez Delgado}, {Robert}, {Kune}, {de Mello}, {Devost}, \&
  {Heckman}}]{1999ApJS..123....3L}
{Leitherer}, C., {Schaerer}, D., {Goldader}, J.~D., {Gonz{\'a}lez Delgado},
  R.~M., {Robert}, C., {Kune}, D.~F., {de Mello}, D.~F., {Devost}, D., \&
  {Heckman}, T.~M. 1999, \apjs, 123, 3

\bibitem[{{Maund} \& {Smartt}(2009)}]{2009Sci...324..486M}
{Maund}, J.~R., \& {Smartt}, S.~J. 2009, Science, 324, 486

\bibitem[{{Mazzali} {et~al.}(2010){Mazzali}, {Maurer}, {Valenti}, {Kotak}, \&
  {Hunter}}]{2010MNRAS.408...87M}
{Mazzali}, P.~A., {Maurer}, I., {Valenti}, S., {Kotak}, R., \& {Hunter}, D.
  2010, \mnras, 408, 87

\bibitem[{{Mortlock} {et~al.}(2009){Mortlock}, {Peiris}, \&
  {Ivezi{\'c}}}]{2009MNRAS.399..699M}
{Mortlock}, D.~J., {Peiris}, H.~V., \& {Ivezi{\'c}}, {\v Z}. 2009, \mnras, 399,
  699

\bibitem[{{Murphy} {et~al.}(2011){Murphy}, {Jennings}, {Williams}, {Dalcanton},
  \& {Dolphin}}]{2011ApJ...742L...4M}
{Murphy}, J.~W., {Jennings}, Z.~G., {Williams}, B., {Dalcanton}, J.~J., \&
  {Dolphin}, A.~E. 2011, \apjl, 742, L4

\bibitem[{{Paragi} {et~al.}(2010){Paragi}, {Taylor}, {Granot}, {Ramirez-Ruiz},
  {Bietenholz}, {van der Horst}, {Pidopryhora}, {van Langevelde}, {Garrett},
  {Szomoru}, {Argo}, {Bourke}, \& {Paczy\'{n}ski}}]{2010Natur.463..516P}
{Paragi}, Z., {Taylor}, G.~B.and~{Kouveliotou}, C., {Granot}, J.,
  {Ramirez-Ruiz}, E., {Bietenholz}, M., {van der Horst}, A.~J., {Pidopryhora},
  Y., {van Langevelde}, H.~J., {Garrett}, M.~A., {Szomoru}, A., {Argo}, M.~K.,
  {Bourke}, S., \& {Paczy\'{n}ski}, B. 2010, \nat, 463, 516

\bibitem[{{Pettini} \& {Pagel}(2004)}]{2004MNRAS.348L..59P}
{Pettini}, M., \& {Pagel}, B.~E.~J. 2004, \mnras, 348, L59

\bibitem[{{Podsiadlowski} {et~al.}(2004){Podsiadlowski}, {Langer},
  {Poelarends}, {Rappaport}, {Heger}, \& {Pfahl}}]{2004ApJ...612.1044P}
{Podsiadlowski}, P., {Langer}, N., {Poelarends}, A.~J.~T., {Rappaport}, S.,
  {Heger}, A., \& {Pfahl}, E. 2004, \apj, 612, 1044

\bibitem[{{Preibisch} {et~al.}(2002){Preibisch}, {Brown}, {Bridges},
  {Guenther}, \& {Zinnecker}}]{2002AJ....124..404P}
{Preibisch}, T., {Brown}, A.~G.~A., {Bridges}, T., {Guenther}, E., \&
  {Zinnecker}, H. 2002, \aj, 124, 404

\bibitem[{{Salpeter}(1955)}]{1955ApJ...121..161S}
{Salpeter}, E.~E. 1955, \apj, 121, 161

\bibitem[{{Sana} {et~al.}(2012){Sana}, {de Mink}, {de Koter}, {Langer},
  {Evans}, {Gieles}, {Gosset}, {Izzard}, {Le Bouquin}, \&
  {Schneider}}]{2012Sci...337..444S}
{Sana}, H., {de Mink}, S.~E., {de Koter}, A., {Langer}, N., {Evans}, C.~J.,
  {Gieles}, M., {Gosset}, E., {Izzard}, R.~G., {Le Bouquin}, J.-B., \&
  {Schneider}, F.~R.~N. 2012, Science, 337, 444

\bibitem[{{Sander} {et~al.}(2012){Sander}, {Hamann}, \&
  {Todt}}]{2012A&A...540A.144S}
{Sander}, A., {Hamann}, W.-R., \& {Todt}, H. 2012, \aap, 540, A144

\bibitem[{{Scheepmaker} {et~al.}(2007){Scheepmaker}, {Haas}, {Gieles},
  {Bastian}, {Larsen}, \& {Lamers}}]{2007A&A...469..925S}
{Scheepmaker}, R.~A., {Haas}, M.~R., {Gieles}, M., {Bastian}, N., {Larsen},
  S.~S., \& {Lamers}, H.~J.~G.~L.~M. 2007, \aap, 469, 925

\bibitem[{{Schmidt} {et~al.}(1994){Schmidt}, {Kirshner}, {Eastman}, {Phillips},
  {Suntzeff}, {Hamuy}, {Maza}, \& {Aviles}}]{1994ApJ...432...42S}
{Schmidt}, B.~P., {Kirshner}, R.~P., {Eastman}, R.~G., {Phillips}, M.~M.,
  {Suntzeff}, N.~B., {Hamuy}, M., {Maza}, J., \& {Aviles}, R. 1994, \apj, 432,
  42

\bibitem[{Schmidt-Kaler(1982)}]{schmidtkaler}
Schmidt-Kaler, T. 1982, in Landolt-B{\"o}rnstein - Group VI Astronomy and
  Astrophysics, Vol.~2b, Stars and Star Clusters, ed. K.~Schaifers \& H.~Voigt
  (Springer Berlin Heidelberg), 14--24

\bibitem[{{Skilling}(2004)}]{2004AIPC..735..395S}
{Skilling}, J. 2004, in American Institute of Physics Conference Series, Vol.
  735, American Institute of Physics Conference Series, ed. {R.~Fischer,
  R.~Preuss, \& U.~V.~Toussaint}, 395--405

\bibitem[{{Smartt}(2015)}]{2015PASA...32...16S}
{Smartt}, S.~J. 2015, \pasa, 32, 16

\bibitem[{{Smartt} {et~al.}(2009){Smartt}, {Eldridge}, {Crockett}, \&
  {Maund}}]{2008arXiv0809.0403S}
{Smartt}, S.~J., {Eldridge}, J.~J., {Crockett}, R.~M., \& {Maund}, J.~R. 2009,
  \mnras, 395, 1409

\bibitem[{{Smith} {et~al.}(2011{\natexlab{a}}){Smith}, {Li}, {Filippenko}, \&
  {Chornock}}]{2011MNRAS.412.1522S}
{Smith}, N., {Li}, W., {Filippenko}, A.~V., \& {Chornock}, R.
  2011{\natexlab{a}}, \mnras, 412, 1522

\bibitem[{{Smith} {et~al.}(2011{\natexlab{b}}){Smith}, {Li}, {Miller},
  {Silverman}, {Filippenko}, {Cuillandre}, {Cooper}, {Matheson}, \& {Van
  Dyk}}]{2011ApJ...732...63S}
{Smith}, N., {Li}, W., {Miller}, A.~A., {Silverman}, J.~M., {Filippenko},
  A.~V., {Cuillandre}, J.-C., {Cooper}, M.~C., {Matheson}, T., \& {Van Dyk},
  S.~D. 2011{\natexlab{b}}, \apj, 732, 63

\bibitem[{{Soderberg} {et~al.}(2010){Soderberg}, {Brunthaler}, {Nakar},
  {Chevalier}, \& {Bietenholz}}]{2010ApJ...725..922S}
{Soderberg}, A.~M., {Brunthaler}, A., {Nakar}, E., {Chevalier}, R.~A., \&
  {Bietenholz}, M.~F. 2010, \apj, 725, 922

\bibitem[{{Valenti} {et~al.}(2008){Valenti}, {Elias-Rosa}, {Taubenberger},
  {Stanishev}, {Agnoletto}, {Sauer}, {Cappellaro}, {Pastorello}, {Benetti},
  {Riffeser}, {Hopp}, {Navasardyan}, {Tsvetkov}, {Lorenzi}, {Patat}, {Turatto},
  {Barbon}, {Ciroi}, {Di Mille}, {Frandsen}, {Fynbo}, {Laursen}, \&
  {Mazzali}}]{2008ApJ...673L.155V}
{Valenti}, S., {Elias-Rosa}, N., {Taubenberger}, S., {Stanishev}, V.,
  {Agnoletto}, I., {Sauer}, D., {Cappellaro}, E., {Pastorello}, A., {Benetti},
  S., {Riffeser}, A., {Hopp}, U., {Navasardyan}, H., {Tsvetkov}, D., {Lorenzi},
  V., {Patat}, F., {Turatto}, M., {Barbon}, R., {Ciroi}, S., {Di Mille}, F.,
  {Frandsen}, S., {Fynbo}, J.~P.~U., {Laursen}, P., \& {Mazzali}, P.~A. 2008,
  \apjl, 673, L155

\bibitem[{{von Hippel} {et~al.}(2006){von Hippel}, {Jefferys}, {Scott},
  {Stein}, {Winget}, {De Gennaro}, {Dam}, \& {Jeffery}}]{2006ApJ...645.1436V}
{von Hippel}, T., {Jefferys}, W.~H., {Scott}, J., {Stein}, N., {Winget}, D.~E.,
  {De Gennaro}, S., {Dam}, A., \& {Jeffery}, E. 2006, \apj, 645, 1436

\bibitem[{{Walmswell} {et~al.}(2013){Walmswell}, {Eldridge}, {Brewer}, \&
  {Tout}}]{2013MNRAS.435.2171W}
{Walmswell}, J.~J., {Eldridge}, J.~J., {Brewer}, B.~J., \& {Tout}, C.~A. 2013,
  \mnras, 435, 2171

\bibitem[{{Williams} {et~al.}(2014){Williams}, {Peterson}, {Murphy}, {Gilbert},
  {Dalcanton}, {Dolphin}, \& {Jennings}}]{2014ApJ...791..105W}
{Williams}, B.~F., {Peterson}, S., {Murphy}, J., {Gilbert}, K., {Dalcanton},
  J.~J., {Dolphin}, A.~E., \& {Jennings}, Z.~G. 2014, \apj, 791, 105

\bibitem[{{Xu} {et~al.}(2011){Xu}, {Nagataki}, \&
  {Huang}}]{2011ApJ...735....3X}
{Xu}, M., {Nagataki}, S., \& {Huang}, Y.~F. 2011, \apj, 735, 3

\bibitem[{{Yoon} {et~al.}(2012){Yoon}, {Gr{\"a}fener}, {Vink}, {Kozyreva}, \&
  {Izzard}}]{2012A&A...544L..11Y}
{Yoon}, S.-C., {Gr{\"a}fener}, G., {Vink}, J.~S., {Kozyreva}, A., \& {Izzard},
  R.~G. 2012, \aap, 544, L11

\end{thebibliography}

\end{document}